\documentclass[journal]{IEEEtran}
\usepackage{array}
\usepackage[numbers]{natbib}
\usepackage{graphicx}
\usepackage{subcaption}
\usepackage{rotating}
\usepackage{natbib}
\usepackage{floatrow}
\usepackage{float} 
\usepackage{tabularx} 
\usepackage{longtable}
\usepackage{booktabs}
\usepackage{acronym}
\usepackage{hyperref}
\usepackage[dvipsnames]{xcolor}

\begin{document}

\title{Deep Learning Techniques in Extreme Weather Events: A Review}

\author{Shikha~Verma,~\IEEEmembership{India Meteorological Department, Ministry of Earth Sciences, New Delhi}\\
        Kuldeep~Srivastava,~\IEEEmembership{India Meteorological Department, Ministry of Earth Sciences, New Delhi}\\
        Akhilesh~Tiwari,~\IEEEmembership{Indian Institute of Information Technology, Allahabad}\\
        Shekhar~Verma,~\IEEEmembership{Indian Institute of Information Technology, Allahabad}\\
}

\maketitle

\begin{abstract}
Extreme weather events pose significant challenges, thereby demanding techniques for accurate analysis and precise forecasting to mitigate its impact. In recent years, deep learning techniques have emerged as a promising approach for weather forecasting and understanding the dynamics of extreme weather events. This review aims to provide a comprehensive overview of the state-of-the-art deep learning in the field. We explore the utilization of deep learning architectures, across various aspects of weather prediction such as thunderstorm, lightning, precipitation, drought, heatwave, cold waves and tropical cyclones. We highlight the potential of deep learning, such as its ability to capture complex patterns and non-linear relationships. Additionally, we discuss the limitations of current approaches and highlight future directions for advancements in the field of meteorology. The insights gained from this systematic review are crucial for the scientific community to make informed decisions and mitigate the impacts of extreme weather events.
\end{abstract}


\IEEEpeerreviewmaketitle

\section{Introduction}
\label{sec:Introduction}

\IEEEPARstart{W}{eather} refers to short-term natural events that occur in a certain location and time which include characteristics such as temperature, pressure, humidity, cloud cover, precipitation, wind speed and wind direction \cite{wmoWorldMeteorological}. Extreme weather, on the other hand, refers to weather events that deviate significantly from the expected conditions. Some instances of extreme weather encompass include tropical cyclones, heatwaves, intense blizzards with heavy snowfall, excessive rainfall leading to flooding, and droughts \cite{easterling2000climate} \cite{beniston2004extreme} \cite{alexander2006global} \cite{tebaldi2006going}. These occurrences pose serious challenges to society and the environment, requiring careful planning and necessary measures to mitigate their detrimental effects. Consequently, predicting weather holds significant importance.

Weather prediction relies on a gathering information from weather stations, satellites, radar systems, weather balloons, and buoys to access the current atmospheric conditions \cite{wmoWorldMeteorological} \cite{imdHomeIndia}. NWP utilizes mathematical models to simulate the pattern of atmosphere which are based on initial conditions collected from observational data \cite{wmoWorldMeteorological} \cite{imdHomeIndia} \cite{espeholt2022deep}. Ensemble forecasting generates several forecasts with slight modifications in initial variables and model parameters to assess uncertainty and likelihood of possible outcomes. Climate models utilize data assimilation, which integrates observational data with model output, to generate long-term weather trends with increased forecast precision.

Deep learning models are composed of multiple layers of interconnected artificial neurons \cite{goodfellow2016deep}. The distinguishing feature of deep learning models is their ability to automatically learn and discover intricate patterns and features directly from the data, without the need for explicit feature engineering. This is achieved by passing the data through multiple layers of interconnected neurons, where each layer learns to extract increasingly abstract representations of the input data. The precision of weather forecasts is heavily dependent on historic data. However, non-linear and complex nature of weather phenomena poses inherent challenges to achieving absolute precision. The traditional methods, including statistical, dynamical and numerical models, have proven effective in forecasting weather events with considerable lead time \cite{bauer2015quiet}, they encounter limitations in capturing intricate patterns and dynamics. As a result, achieving accurate predictions becomes unattainable due to the intricate nature of weather systems. Continuous advancements in research and the utilization of emerging technologies, such as deep learning, offer promising avenues for further enhancing weather prediction capabilities. This transformative approach enables highly accurate weather predictions including severe weather events, empowering proactive measures to mitigate their impacts effectively. Deep learning facilitates the integration of diverse data sources, including satellites, radars, and weather stations, to provide comprehensive and real-time meteorological insights for improved public safety and resilience.\\

This review is organized as follows: \autoref{sec:Introduction}  introduces the paper by providing an overview of the challenges, advancements, and applications of weather forecasting using deep learning, as well as outlining the organization of the paper. 
\autoref{sec:Extreme_Weather} explores the realm of extreme weather events, highlighting the need of accurate weather prediction. 
In \autoref{sec:Deep_Learning}, an extensive literature review is presented, focusing on the utilization of deep learning for extreme weather events. This section explores the existing research, highlighting the different approaches, models, and findings in the field. 
\autoref{sec:challenges} focuses on the challenges encountered in the field of deep learning for meteorology and weather forecasting. This section discusses the limitations, data issues, interpretability concerns, and other obstacles faced when applying deep learning techniques in this domain. 
\autoref{sec:discussion} highlights how the integration of deep learning in extreme weather research advances predictive capabilities and emphasizes the potential of hybrid models to enhance forecast accuracy and proactive mitigation strategies. Additionally, it outlines potential avenues for future research in weather forecasting using deep learning, discussing promising areas of exploration, methodologies, and potential advancements to enhance the accuracy and efficiency of deep learning models for weather prediction.
Finally, \autoref{sec:conclusion} offers a concise conclusion that summarizes the key findings and contributions of the study. 

\section{Extreme Weather Events}
\label{sec:Extreme_Weather}
Exploring the fundamental elements that govern weather patterns and shape the dynamics of atmosphere helps to obtain a broader perspective \cite{wmoWorldMeteorological}. Temperature plays a pivotal role in determining the thermal state of the atmosphere, thereby exerting a substantial influence over the behaviors of gases, liquids, and the overall human comfort. 
Air pressure, commonly referred to as atmospheric pressure, significantly shapes weather patterns by creating low and high-pressure systems. These systems contribute to a wide array of meteorological conditions, including notable adverse conditions in case of low-pressure systems.
The pressure difference give rise to atmospheric winds that traverse from low pressure area to high pressure area, enabling the essential mechanism of air circulation. 
Humidity, the amount of water vapour present in atmosphere, demonstrates a direct relationship with temperature; higher temperatures allow the air to hold more water vapor, leading to the formation of clouds. 
The extent of cloud cover plays a decisive role in modulating solar radiation reaching the Earth's surface, thereby affecting temperature profile and atmospheric dynamics. 
Precipitation, which includes rain, snow, and hail, occurs when moisture in the atmosphere condenses and is then released from clouds. This process constitutes a primary mechanism through which atmospheric water is returned to the Earth's surface \cite{trenberth2011changes}
When these atmospheric elements surpass anticipated norms, they can trigger a diverse range of extreme weather events that carry significant consequences. These include heatwaves posing health risks and intensifying wildfires \cite{abdin_modeling_2019}, intense precipitation leading to flooding and landslides, warm ocean conditions fueling cyclones with their devastating winds and storm surges, severe thunderstorms generating tornadoes and hailstorms, prolonged droughts affecting agriculture and water supply, snowstorms and blizzards disrupting transportation and infrastructure, and freezing rain causing damaging ice storms. Monitoring and understanding these factors are essential for early detection, preparedness, and effective management to mitigate the potential impacts of these diverse extreme weather phenomena \cite{coumou_decade_2012}. 
To achieve this, ensemble forecasting emerges as a powerful tool, generating several forecasts with slight modifications in initial variables and model parameters to assess uncertainty and the likelihood of possible outcomes. This strategic approach is complemented by the utilization of climate models, which employ data assimilation techniques. By integrating observational data with model outputs, these models generate long-term weather trends with precise forecast. These forecasts enable decision-makers and emergency response teams to better understand the level of risk associated with different weather events and make informed choices to protect communities and infrastructure. 

\section{Deep Learning in Extreme Weather Events}
\label{sec:Deep_Learning}
Deep learning models are revolutionizing extreme weather prediction by leveraging diverse data sources to accurately forecast events such as cyclones, heatwaves, heavy rainfall, and severe storms. Their ability to analyze complex patterns and relationships enables early warning systems and proactive mitigation strategies, with the potential to minimize the impacts of extreme weather on society and the environment. 

A Deep Neural Network (DNN) is being developed for an early warning system to predict extreme weather events such as floods, droughts, and heatwaves. The DNN approach effectively downscales and bias corrects coarse resolution seasonal forecast ensembles, generating realistic, high-resolution climate information. The study demonstrates that the DNN model accurately predicts extreme values while preserving the physical relationships and trends in the variables \cite{heidari_towards_2023}. Researchers aim to improve the forecast of SCW, including thunderstorms, short-duration heavy rain, hail, and convective gusts by employing deep-CNN algorithm to effectively extract the characteristics of SCW and achieve better forecast performance as compared to traditional machine learning algorithms \cite{zhou_forecasting_2019}. Deep learning methods are being proposed for detecting and forecasting anomalies in spatiotemporal data. The choice of learning task depends on whether the anomalies are known or unknown. Anomalies are often imbalanced and require specific data pre-processing. Leveraging diverse data sources and modelling techniques, deep learning models exhibit strong capabilities in accurately forecasting flood occurrence, severity, and spatial distribution. These models play a crucial role in real-time monitoring, enabling timely response and effective mitigation strategies. Wasserstein Generative Adversarial Network (WGAN) is being utlized for downscaling tropical cyclone rainfall to hazard-relevant spatial scales \cite{vosper_deep_2023}. Additionally, a hybrid approach combining WGAN and Variational Autoencoder GAN (VAEGAN) is being introduced to enhance the resolution of rainfall measurements from 100 km to 10 km resolution, showing realistic power spectra for various wave numbers \cite{vosper2023deep}. A deep learning-based technique employing a fully connected neural network is being proposed to accurately predict rainfall-induced shallow landslides across Italy \cite{mondini_deep_2023}.

\subsection{Deep Learning in Thunderstorm and Lightning}

Thunderstorms are complex atmospheric phenomena characterized by a combination of thunder, lightning, heavy rainfall, and strong winds, with lightning resulting from electrical discharges within clouds or between clouds and the ground.

\begin{table*}[htp]
\caption{Deep Learning in Thunderstorm and Lightning}
\label{}
\centering
\begin{tabular}{@{}llll@{}}
\toprule
Task & Approach & Ref\\ 
\midrule

Thunderstorm Prediction & EEMD-ANN, EEMD-SVM, ARIMA
& \citep{azad_development_2021}\\

& LRCN-CNN, LSTM
& \citep{guastavino_prediction_2022}\\

Thunderstorm Severity Prediction & LSTM-FC, CNN-LSTM, ConvLSTM
& \citep{essa_deep_2022}\\

\midrule

Lightning Prediction & RNN
& \citep{lin_attention-based_2019}\\

& ResNet
& \citep{lu_monitoring_2022}\\

Lightning Identification & CNN
& \citep{qian_lightning_2022}\\

\bottomrule
\end{tabular}
\end{table*}

In a study two hybrid models, EEMD-ANN and EEMD-SVM, are being developed for predicting thunderstorm frequency in Bangladesh. These models utilize ensemble empirical mode decomposition (EEMD) to extract relevant features for accurate prediction. The EEMD-ANN model consists of an input layer with 11 variables, two hidden layers (4 and 2 neurons), sigmoid activation function, and a 0.1 learning rate. On the other hand, EEMD-SVM employs various kernel functions for effective handling of non-stationary TSF data. The inpur variables include CAPE, CPRCP, CRR, DP, KI, PRCP, RH, ST, TSD, TT, and WS50. EEMD, based on Hilbert-Huang transform, mitigates challenges of EMD by introducing Gaussian white noise. This enables precise decomposition of time series data into intrinsic mode functions (IMFs), revealing underlying patterns. ARIMA models effectively handle non-stationary time series data. The hybrid models, EEMD-ANN and EEMD-SVM, capitalize on EEMD's capabilities to handle non-stationary data and capture nonlinear relationships. These models outperform like ANN, SVM, and ARIMA in terms of prediction accuracy, with improvements ranging from 8.02\% to 22.48\% across TSF categories \cite{azad_development_2021}. 
In another study, the aim is to predict severe thunderstorm occurrences through an innovative approach using lightning and radar video data in the Liguria region of Italy. The ensemble technique outperforms traditional methods that optimized standard quality-based scores. The architecture involves a LRCN, which combines a CNN for extracting spatial features and an LSTM network for analyzing sequential aspects. The training process spans 100 epochs, employing the Adam Optimizer with a learning rate of 0.001 and a mini-batch size of 72. The training process incorporates a class balanced cross-entropy loss function to fine-tune the model's performance. The model's reliability is validated using a historical radar video dataset comprising CAPPI images at 2 km, 3 km, and 5 km above above sea level, demonstrating its effectiveness in probabilistic forecasting of severe thunderstorms \cite{guastavino_prediction_2022}. 
In a study focused at assessing the accuracy of various LSTM neural network variants in predicting thunderstorm severity through the utilization of remote sensing weather data, the primary objective is to quantitatively forecast the intensity of thunderstorms by analyzing the frequency of lightning flashes using deep learning models. The study employs two main datasets: SALDN lightning detection network data and SAWS weather station data. These datasets are used to train and evaluate different LSTM neural network variants, including LSTM-FC, CNN-LSTM, and ConvLSTM models. The LSTM-FC model consists of three LSTM layers and one dense layer, with an optimizer based on the Adam algorithm. The activation function used is the Leaky-Rectified Linear Unit with an alpha value of 0.15. Similarly, the CNN-LSTM model comprises two Conv2D layers, one LSTM layer, and one dense layer. This model also employs the Adam optimizer and utilizes the Leaky-Rectified Linear Unit as the activation function, with an alpha value of 0.05. The ConvLSTM model is structured with two ConvLSTM2D layers and two dense layers. The same Adam optimizer is employed, and the activation functions are set to Leaky-Rectified Linear Unit (with an alpha value of 0.05) and Rectified Linear Unit. The models are trained and evaluated using hourly lightning flash data and weather variables based on the MAE and MSE. Among the various LSTM model variants, the CNN-LSTM model outperforms the other models with a MAE of 51 flashes per hour because of its ability to capture spatio-temporal features, leading to more accurate predictions of thunderstorm severity \cite{essa_deep_2022}. 

To predict the occurrence of lightning, an innovative data-driven neural network model called Attention-Based Dual-Source Spatiotemporal Neural Network (ADSNet) is being introduced. ADSNet is designed for accurate hourly lightning forecasting and utilizes both numerical simulations and historical lightning observations, resulting in a comprehensive and effective approach. A diverse dataset, combining WRF simulation data with Cloud-to-Ground Lightning Location System (CGLLS) observations from North China, is being employed. The model consists of dual RNN encoder-decoder units, several CNN modules, DCNN modules, and attention mechanisms. ConvLSTM is chosen for its adeptness in capturing intricate spatiotemporal dependencies. This intricate framework is tailored for conducting 12-hour lightning forecasts in the North China region. The model adopts the Adam optimizer with an initial learning rate of 0.0001 and Weighted Binary Cross-Entropy as a loss function. Experimental results are validating the superiority of ADSNet over baseline methods in terms of lightning forecast accuracy \cite{lin_attention-based_2019}. 
In another study, an innovative approach known as Lightning Monitoring Residual Network (LM-ResNet) is being introduced, leveraging deep learning for effective lightning location monitoring in Ningbo, China. By transforming the task into binary classification, radar data (PPI, CR, ET, V) and essential land attributes (DEM, aspect, slope, land use, NDVI) are being harnessed to create a comprehensive lightning feature dataset. LM-ResNet employs Rectified Linear Unit (ReLU) activation for effective learning and addresses data imbalances through Focal Loss, a specialized cross-entropy-based loss function. The model's training configuration includes an initial learning rate of 0.1 and utilizes the SGD optimizer with a batch size of 64, incorporating momentum of 0.9 and weight attenuation of 0.0004 to enhance learning while mitigating overfitting. The study is demonstrating LM-ResNet's superiority over competing architectures like GoogLeNet and DenseNet, highlighting its potential for accurate and reliable lightning incident tracking. \cite{lu_monitoring_2022}. 

An approach called Lightning-SN is introduced, designed for precise cloud-to-ground (CG) lightning identification using deep learning techniques. This model effectively utilizes S-band Doppler radar data and CG lightning records of the Ningbo area in Zhejiang Province, China, collected from August 2009 to December 2021 via the ADTD lightning positioning system. Lightning-SN leverages an encoder-decoder structure with 25 convolutional layers, five pooling layers, five upsampling layers, and a sigmoid activation function layer. The architecture capitalizes on symmetry, boundary preservation techniques, and a 1x1 convolution kernel in the final layer. The model's optimization is driven by the Adam optimizer and guided by the GHM loss function. Training involves the BP algorithm, employing iterative refinement and validation testing. Additionally, the study includes a comprehensive comparative analysis with other semantic segmentation algorithms—FCNN, DeepLab-V3, and BiSeNet—evaluated under identical conditions. Lightning-SN demonstrates substantial performance improvements over traditional threshold-based methods, particularly in scenarios involving high-resolution radar data. \cite{qian_lightning_2022}.

\subsection{Deep Learning in Precipitation}
Precipitation is the process by which water, in either liquid or solid form, falls from the atmosphere to the Earth's surface. Hail, snow, and rainfall are the three most common types of precipitation. The most frequent type of precipitation is rain, which occurs when water droplets congregate and become heavy enough to fall to the ground. When raindrops are pushed higher into the freezing parts of the sky during violent thunderstorms, they freeze and pile in layers, resulting in hailstones of varied sizes that can cause property and crop damage. Snow is formed when water vapour condenses straight into ice crystals in cold atmospheric conditions. Accurate forecasting and understanding of precipitation patterns are critical for many industries, including agriculture, water resource management, and transportation. \\

\begin{table*}[htp]
\caption{Deep Learning in Precipitation}\label{table_precipitation}
\centering
\begin{tabular}{@{}llll@{}}
\toprule
Task & Approach & Ref\\ 
\midrule

Precipitation Forecast 
& RNN
& \citep{yao_improved_2022}\\

& CNN
& \citep{espeholt_deep_2022}\\

Precipitation Data Merging & CNN, ANN
& \citep{nan_deep_2023}\\

Quantitative Precipitation Estimation & CNN-based
& \citep{li_polarimetric_2023}\\

TPW and CAPE Estimation & MLP
& \citep{lee_incremental_2022}\\

\midrule

Hailstorm Detection & CNN, DNN
& \citep{pullman_applying_2019}\\

Hailstorm Forecast & Atuencoder, CNN
& \citep{pulukool_using_2020}\\

& CNN
& \citep{ii_interpretable_2019}\\

Hail Size Estimation & PCA, BPNN
& \citep{wu_estimation_2022}\\

\midrule

Cloud or Snow Identification 
& DeepLab-CRF
& \citep{wang_cloud_2022}\\

& CNN 
& \citep{zhan_distinguishing_2017}\\

& U-Net
& \citep{yin_cloud_2022}\\
 
& U-Net
& \citep{sood_glacier_2022}\\

& U-Net
& \citep{wang_snow_2022}\\

& CNN
& \citep{wang_automated_2021}\\

Snow Depth Estimation 
& BPNN
& \citep{zhu_downscaling_2021}\\

& CNN, ResNet
& \citep{xing_estimation_2022}\\

& deep CNN
& \citep{yao_snow_2022}\\

Snow Water Equivalent Estimation & ANN, ANFIS
& \citep{ghanjkhanlo_prediction_2020}\\

& MNLR, NNGA
& \citep{marofi_predicting_2011}\\

\bottomrule
\end{tabular}
\end{table*}

\subsubsection{Rainfall}
Deep learning uses meteorological data such as historical rainfall records, satellite images, and atmospheric conditions. Rainfall forecasts give essential information for disaster preparedness, agricultural planning, water resource management, and climate modelling. 

A nowcasting model is designed to address extreme weather phenomena encompassing both precipitation and landfalling hurricanes. The research employs a comprehensive dataset spanning five years (2015–2020) of radar observations over South Texas, including 22 hurricane events that occurred in the United States. The architectural design of the model comprises four core components: RNN, up-sample, down-sample, and convolution. The architecture is built upon a three-layer encoder-decoder structure, incorporating distinct filter arrangements for the RNN, while seamlessly integrating convolution and deconvolution operations. GRU is selected as the foundational RNN unit, organised in multiple layers to effectively capture intricate spatiotemporal patterns. The model effectively predicted future radar reflectivity echo maps based on five preceding observations, enabling forecasts for up to a 3-hour lead time. Model parameter optimization is achieved using the Adam optimizer, fine-tuned with a learning rate of 10\textsuperscript{4} and a momentum of 0.5. To further enhance predictive performance, the research incorporates Balanced Mean Squared Error (B-MSE) and Balanced Mean Absolute Error (B-MAE) as loss functions. The model's forecasting capabilities are evaluated using established metrics—HSS, CSI, POD, and FAR—all of which collectively highlight its proficiency in precipitation nowcasting. \cite{yao_improved_2022}. 
MetNet-2, a deep neural network-based weather model, outperforms existing physics-based models in predicting high-resolution precipitation up to 12 hours ahead. The study utilizes data sources such as MRMS, GOES-2, and HRRR datasets. Input observations from various sources, including radar, satellite, and assimilation features, are processed through a CNN to capture temporal dynamics. Efficient computation is achieved through model parallelism across 16 interconnected TPU cores, allowing accurate forecasts over a 512 km x 512 km target patch. The model's architecture consists of three stacks of 8 residual blocks with exponentially increasing dilation factors. Operating within the Continental United States, MetNet-2 generates forecasts at a 2-minute frequency with a spatial resolution of 1 km. It operates within a probabilistic framework, producing categorical predictions across 512 precipitation levels for each target position. The model's performance exceeds that of the High-Resolution Ensemble Forecast (HREF) when assessed using the Cumulative Ranked Probability Score (CRPS). \cite{espeholt_deep_2022}.

The utilization of deep learning techniques to merge precipitation data from diverse sources across the Tibetan Plateau, with the aim of enhancing data precision. The study explores three methodologies: ANN, CNN and a statistical Extended Triple Collocation (ETC) method. The neural network architecture employed consists of an ANN with four fully connected layers and a CNN enhanced by two additional convolutional layers to capture spatial features. To mitigate overfitting, dropout layers with a 0.1 dropout rate followed each fully connected or convolutional layer. The optimization employs the Adam algorithm with a learning rate of 0.0001, the RMSE serves as the loss function, and the ReLU function acts as the activation function. The hyperparameters consist of 500 epochs and a batch size of 2500 for effective training. Meteorological and hydrological evaluations reveal that the CNN approach consistently demonstrates superior performance compared to the others, showcasing enhanced spatial distribution and heightened accuracy. The meteorological evaluation employs eight metrics: CC, BIAS, STDRATIO, MAE, RMSE, POD, FAR, and CSI. The hydrological assessment utilizes NSE and PBIAS for model parameter validation, with KGE employed to counter NSE's flow peak bias and emphasize runoff variability \cite{nan_deep_2023}.

Using the U.S. Weather Surveillance Radar-1988 Doppler (WSR-88D) observations dataset, researchers developed four DL models for QPE (Quantitative Precipitation Estimation) with a CNN-VGG architecture. These models, named RQPENetD1, RQPENetD2, RQPENetV, and RQPENetR, incorporate dense blocks, RepVGG blocks, and residual blocks. The architecture of RQPENetD1 features an initial convolution layer, four dense blocks with varying bottleneck layers, and transition layers for spatial reduction. It processes 3-D radar data from two elevation angles to estimate rainfall rate using a fully connected layer with adaptive average pooling and utilizes MSE as the loss function. RQPENetD2 shares a similar structure with dense blocks featuring (24, 16) and (36, 24) bottleneck layers, along with transition layers involving 1x1 Convolution and average pooling. RQPENetV incorporates RepVGG blocks in a multi-branch structure across five stages, while RQPENetR utilizes residual modules in four sequential blocks with varying bottleneck layers for feature enhancement from 3-D radar data. The evaluation of RQPENet's radar precipitation estimation includes metrics such as RMSE, MAE, CC, and NSE, along with additional atmospheric science metrics: POD, FAR, CSI, HSS, and GSS. The findings indicate the superior performance of dense blocks-based models, particularly RQPENet D1 and RQPENet D2, compared to residual blocks and RepVGG blocks-based models, as well as five traditional Z-R relations \cite{li_polarimetric_2023}.

In a recent study, researchers propose an ANN model with incremental learning to derive total precipitable water (TPW) and convective available potential energy (CAPE) from GEO-KOMPSAT-2A satellite imagery over Northeast Asia. The study utilizes AMI satellite imagery, ERA5 data, and radiosonde observations for training and evaluation. An MLP feedforward backpropagation ANN model is employed for the retrieval algorithm. The model architecture includes an input layer with 20 neurons, a hidden layer with 40 neurons using a hyperbolic tangent activation function, and an output layer with a linear activation function. The optimization process utilizes the Adam optimizer with a mean squared error loss function. The accuracy assessment involves the utilization of statistical metrics, including correlation coefficient, bias, and RMSE. The incremental ANN model demonstrates improved accuracy and stability compared to static learning methods, indicating its potential to accurately estimate TPW and CAPE \cite{lee_incremental_2022}. \\

\subsubsection{Hail}
Hail prediction helps to improve our understanding and readiness for this dangerous weather occurrence.  We can increase hail forecasting accuracy, giving advanced warnings to prevent possible infrastructure, agriculture, and community damage. Because hail storms may have significant socioeconomic consequences, applying deep learning techniques into hail prediction models is critical for timely and effective risk management and disaster response measures. 

In a test case study, researchers applied deep learning networks for hailstorm detection using CNN and DNN architectures. The approach involves training these networks on GOES satellite imagery and MERRA-2 atmospheric parameters to identify hail storms. Different architectures are utilized, including a CNN for processing satellite imagery and a DNN for atmospheric parameters, aimed at capturing pertinent features. The CNN architecture for satellite imagery uses four convolutional layers with ReLU activation functions, combined with max-pooling layers for downsizing and a batch normalization layer for streamlined training. Fully connected layers are also integrated into the architecture to enable classification. Concurrently, the DNN architecture for atmospheric parameters features four fully connected layers with ReLU activation, utilizing a Softmax function for classification. Both architectures converge within a merged network, amalgamating outputs from the CNN and DNN via concatenation and incorporating additional fully connected layers for the final classification. This approach harnesses the capabilities of deep learning to enhance hail detection by merging multi-source data and recognizing spatial patterns. The CNN model achieves heightened precision by accurately identifying decreased infrared brightness temperatures linked to hail storms. \cite{pullman_applying_2019}.
In an effort to forecast hailstorms, researchers introduced an architecture that comprises three distinct models: an Autoencoder (AE) with encoder and decoder layers, each containing 32 neurons; a CNN constructed with CNN layers featuring 64 and 32 filters; and an RF model characterized by an ensemble of decision trees and decision tree aggregation through majority voting. Both AE and CNN are optimized using the Adam optimizer and MSE as the loss function. The dataset utilized in the study consists of observations from the TRMM and reanalysis data from the ECMWF spanning one year. The selected attributes for training the models include convective potential energy, convective inhibition, wind shear within the 1–3 km range, and warm cloud depth. The study aims to predict global hailstorms using these models and to compare their performance in terms of accuracy, precision, and error rates. Surprisingly, contrary to expectations, RF outperforms the deep learning methods in terms of hailstorm prediction performance \cite{pulukool_using_2020}. 
In a study focusing on severe hail prediction, researchers employed CNN to encode spatial weather data and compared its performance with traditional statistical approaches like Logistic Mean and Logistic PCA. The dataset utilized in this study includes geopotential height, temperature, dewpoint, zonal wind, and meridional wind variables from the NCAR ensemble model output. These variables are collected at different pressure levels: 500 hPa, 700 hPa, and 800 hPa. The study uses upper-air dynamic and thermodynamic fields from an NCAR NWP model. The CNN architecture used in this study comprises three strided convolutional layers with 5x5 gridcell filters. A range of hyperparameters is tested, including the initial number of filters, dropout rates, activation functions (ReLU and Leaky ReLU), L2 norm regularization coefficients, and optimizers (Stochastic Gradient Descent and Adam) with different learning rates. The model's evaluation is conducted using the Brier Score as the prediction error function, and standard probabilistic verification metrics are employed to assess the quality of probabilistic forecasts. The results demonstrate a significant enhancement in various measures of prediction skill achieved by the CNN architecture, leading to improved probabilistic predictions when compared to the logistic PCA approach \cite{ii_interpretable_2019}.

Accurately estimating hail size remains crucial for evaluating potential damage caused by hailstorms. In order to address this challenge, a model consisting of two main components has been proposed. The PCA-based technique selects 18 features that strongly correlate with hail sizes, while the BPNN regression model with a two-layer architecture and 35 hidden layer neurons is employed to estimate the size of hailstones from satellite images. Using a MSE loss function, the BPNN regression model achieves an R-squared value of 0.52 through linear fitting when assessing the correspondence between predicted and observed Maximum Hail Diameters on the test set \cite{wu_estimation_2022}. \\

\subsubsection{Snow}
Understanding snowfall patterns is critical for many industries, including transportation, agriculture, and disaster planning, resulting to more effective resource management and risk mitigation techniques. However, the intricate task of accurately differentiating between snow and clouds arises from their comparable white appearance in satellite imagery, highlighting the utmost importance of precise discrimination. 
Therefore, in an attempt to accurately identify cloud and snow in high-resolution remote sensing images, a study introduced the DeepLab v3+ neural network with a CRF model. The research utilized data from China Gaofen-1 (GF-1) satellite's Wide Field View (WFV) sensor, comprising four bands and a spatial resolution of 16 m, encompassing a total of ten images spanning three years. The DeepLab v3+ adopts an encode-decoder architecture and employs the Adam optimizer with a learning rate of 0.001, a batch size of 5, and 200 epochs. The study analyzes accuracy variations resulting from distinct loss functions, including the Cross Entropy (CE) loss function, Dice loss function, and Focal loss function. Evaluation metrics encompass Mean Intersection over Union (MIoU) and Mean Pixel Accuracy (MPA) to assess model performance. This methodology effectively mitigates misclassification issues, enhancing cloud and snow identification precision through refined boundary delineation and reduced isolated patches \cite{wang_cloud_2022}. 
An end-to-end fully-CNN with a multiscale prediction approach is proposed to differentiate cloud and snow using a dataset of 50 high-resolution Gaofen satellite images (13400×12000 pixels each), meticulously labeled for cloud and snow regions. The network adopts the VGG network architecture with stride reduction and atrous convolution techniques. Due to the frequent co-occurrence of snow and cloud in images, a pixel-level approach is employed, involving the replacement of the last two fully connected layers in the VGG model with two convolutional layers. The final layer employs a three-class softmax loss for classifying snow, cloud, and other land types, using batch normalization and rectified linear units. The Multiscale Prediction Module merges feature maps from diverse intermediate layers, functioning as an ensemble learning approach. This approach allows for the simultaneous utilization of low-level spatial information and high-level semantic information, enabling accurate differentiation between cloud and snow \cite{zhan_distinguishing_2017}. 
A deep learning-based method is developed by utilizing the Unet3+ network with Resnet50 and the Convolutional Block Attention Module (CBAM) to accurately detect cloud and snow in remote sensing images. This approach effectively eliminates interference information. The feature extraction process of UNet3+ includes five encoders with effective convolutional and pooling layers. In an enhanced version, multiple convolutional layers, regularization, ReLU activation, and residual modules are added to each of the five encoders, resulting in feature graphs. The decoders consist of convolution and activation layers. To address bias, a weighted cross-entropy loss is employed, emphasizing cloud and snow regions. For enhanced focus and deeper feature extraction, the Convolutional Block Attention Module (CBAM) is incorporated into ResNet50. CBAM harnesses channel attention through global pooling, multi-layer perceptron processing, and sigmoid activation to generate attention feature maps. The model's performance is assessed using various metrics, including Mean Intersection over Union (mIoU), Mean Pixel Accuracy (mPA), Mean Precision (mPrecision), and Estimated Total Size. These evaluations successfully mitigate interference data, resulting in accurate cloud and snow extraction from diverse landforms within remote sensing images.\cite{yin_cloud_2022}.
The effectiveness of U-Net based deep learning models in delineating glacier boundaries and identifying snow/ice is demonstrated in a study that developed ENVINet5 and ENVI Net-Multi deep learning classifiers to analyze Landsat-8 satellite data over the Bara Shigri glacier region in Himachal Pradesh, India. The ENVINet5 architecture, based on a mask-based encoder-decoder U-Net model, is employed for single-class categorization, while ENVI Net-Multi is used for multi-class classification of features like snow, ice, and barren areas. The ENVINet5 architecture comprises five levels with twenty-three convolutional layers, incorporating input patches, feature maps, various convolutions, feature fusion, max-pooling, co-convolution, and 1x1 convolutions. For ENVINet-Multi, training parameters include 25 epochs, a patch sampling rate of 16, class weight of 2.5, loss weight of 0.5, 200 patches per epoch, 464x464 pixel patch size, and 2 patches per batch \cite{sood_glacier_2022}.

The performance of U-Net, RF, and Sen2Cor models for snow coverage mapping is compared in a study using Sentinel-2 satellite multispectral images across 40 diverse sites spanning all continents except Antarctica. A Random Forest model is built using Bayesian Hyperparameter Optimization for improved performance. The Sentinel-2 Level-2A product incorporates cloud and snow confidence masks derived from Sen2Cor, which employs threshold tests on spectral bands, ratios, and indices like NDVI and NDSI. A U-Net network architecture is employed, featuring an encoding path with repeated 3x3 convolutions, batch normalization, and ReLU activation, followed by 2x2 max pooling for downsampling. The decoding path utilizes transpose convolutions for upsampling, concatenating with corresponding encoding path features, and applying 3x3 convolutions with BN and ReLU. The final layer comprises a 1x1 convolution for class prediction. Training involves weighted cross-entropy loss, stochastic gradient descent with a learning rate of 0.01 and momentum of 0.9, resulting in effective semantic segmentation. The model's performance is assessed using precision, recall, F1 score, Intersection Over Union (IoU), and accuracy metrics. The results demonstrate that U-Net models exhibit superior performance compared to RF and Sen2Cor in accurately mapping snow coverage \cite{wang_snow_2022}.
An open-source machine learning-based system for snow mapping, AutoSMILE, was developed to automate the process using image processing, machine learning, deep learning, and visual inspection. It was applied in a mountainous area in the northern Tibetan Plateau using RF and CNN algorithms, achieving accurate snow cover mapping. The CNN architecture comprises four layers: convolutional layers for feature extraction, activation layers like ReLU to expedite training, pooling layers for non-linear downsampling, and fundamental components like fully connected and flatten layers. For model evaluation, key metrics include producer's accuracy (PA), user's accuracy (UA), intersection over union (IoU), and overall accuracy (OA) \cite{wang_automated_2021}.

A deep learning approach is introduced to downscale snow depth retrieval across an alpine region by integrating satellite remote-sensing data with diverse spatial scales and characteristics. The study focuses on collaborative snow parameter retrieval in Northern Xinjiang, China, utilizing MODIS and MWRI data. A three-hidden-layer neural network is designed with 20, 20, and 10 neurons in each layer. The network processes resampled BTD, topographic, and meteorological data at a 500m resolution, utilizing a sigmoid function for capturing nonlinear patterns. Backpropagation, guided by MSE and SGD, optimizes the network to enhance the accuracy of snow depth observations. The optimization process entails weight and bias adjustments, informed by a learning rate of 0.001. This approach aims to boost the precision of snow depth observations through deep neural network downscaling. Reference data from ground station snow depth measurements are employed to evaluate the downscaling model's performance and retrieval accuracy, employing assessment metrics encompassing R2, RMSE, PME, NME, MAE, and BIAS  \cite{zhu_downscaling_2021}. 

A deep learning model is presented for 'area-to-point' snow depth estimation, which integrates AMSR2 TB, MODIS, and NDSI data, achieving high accuracy with a spatial resolution of 0.005°. The model utilizes CNN and residual blocks to capture spatial heterogeneity and leverage high-resolution snow information from MODIS. The CNN comprises convolutional and ReLU activation layers, pooling for downsampling, measures to prevent overfitting, and concludes with a fully connected layer for output. The proposed deep residual network contains a 35-layer input patch, applies convolutions with batch normalization and max pooling, followed by 4 residual blocks for feature extraction. After adaptive average pooling and fully connected layers, it predicts snow depth at the patch center. It has 9 convolutions and 4 fully connected layers, using ReLU activation except for the linear output layer. The model is trained for 50 epochs, using a learning rate of 0.0001, with exponential decay of 0.5 every 20 epochs, and a batch size of 32, while employing stochastic gradient descent (SGD) as the optimizer. The evaluation metrics in this study include RMSE, MAE, MBE, and R2. The results demonstrate that by incorporating spatial heterogeneity and leveraging high-resolution MODIS snow cover data, the proposed model achieves promising accuracy in snow depth estimation, with potential applicability to other regions \cite{xing_estimation_2022}. 

A novel inverse method is presented for extracting snow layer thickness and temperature from passive microwave remote sensing data. Utilizing convolutional, pooling, and fully-connected layers, the study employs a ConvNet to inversely estimate the thickness and temperature of a snowpack from its corresponding vertical polarization brightness temperature and horizontal polarization brightness temperature. The model chooses the Adam optimizer with a learning rate of 0.01 to optimize the half mean squared error loss function, and L2 regularization is employed to enhance prediction accuracy by mitigating over-fitting. Furthermore, a comparative analysis is conducted between the ConvNet outcomes and those of conventional ANN and SVM. The model assessment is carried out using RMSE and R2 metrics, underscoring the effectiveness of the advanced ConvNet approach. The utilized ANN architecture comprises three layers - input, hidden, and output - with 20 hidden-layer units utilizing hyperbolic tangent basis functions \cite{yao_snow_2022}. 

A study on predicting snow water equivalent (SWE) in a semi-arid region of Iran was conducted using regression, ANN, and adaptive neuro-fuzzy inference system (ANFIS) models. The study proposes a three-layer ANN alongside ANFIS, which integrates ANN with fuzzy logic, utilizing a five-layer structure based on the Sugeno model featuring two fuzzy if-then rules. In the ANFIS architecture, a hyperbolic tangent activation function is utilized in the hidden layer, and optimal neuron counts are ascertained for both hidden and input layers through iterative refinement. The handling of numerous independent input variables in the ANFIS approach is accomplished using backpropagation training and a sub-clustering method. The assessment of ANN, ANFIS, and regression models involves statistical metrics such as MBE, MAE, RMSE, correlation coefficient, relative error percentage, and Nash-Sutcliffe coefficient efficiency. The results demonstrate the superior performance of both ANN and ANFIS models compared to the regression method, with ANN and ANFIS exhibiting similar prediction accuracy for SWE \cite{ghanjkhanlo_prediction_2020}. 
In a study, researchers focused on estimating SWE in the Samsami basin of Iran using MNLR, NNGA, and ANN architectures. The study aims to estimate snow water equivalent, a critical component of water resources in mountainous areas, based on climatic and topographic parameters such as elevation, slope, aspect, longitude, and latitude. The MNLR architecture employed in the study aims to model the complex non-linear relationship between SWE and a set of independent parameters. Moreover, four different ANN architectures are investigated: MLP for supervised prediction with input, hidden, and output layers; GFF for efficient problem-solving through multi-layer connections; RBF for rapid learning via self-organizing hidden layers; and MNN, a specialized MLP with parallel sub-modules for specialized function and faster training. The study evaluates these architectures to enhance the prediction of SWE values. The NNGA model utilizes genetic algorithms to optimize neural network parameters, enhancing accuracy by iteratively refining weights through selection, crossover, and mutation. Additionally, six diverse learning algorithms are examined for training neural network components in the NNGA model. These algorithms include Levenberg-Marquardt for adaptive MSE minimization, Delta-Bar-Delta for efficient step size adaptation, Step for gradient descent with step size adjustment, Momentum for inertia-infused gradient descent, Conjugate Gradient for second-order optimization, and Quickprop for error surface curvature-based weight adjustments. The study evaluates three activation functions: Sigmoid, Tanh, and Linear, and assesses all models using standard statistical criteria, including correlation coefficient, RSME, ratio of average estimated to observed values, and MAE. The NNGA model, specifically the NNGA5 variant, proves to be the most effective approach, offering valuable insights for water resource management in mountainous regions \cite{marofi_predicting_2011}.

\subsection{Deep Learning in Drought}
Drought refers to an extended period of unusually low precipitation within the natural climate cycle caused by deficiency in rainfall. It can have far-reaching impacts, including water shortages, crop failure, livestock losses, increased wildfire risk, and ecosystem degradation. Drought can be categorized into four types: meteorological, hydrological, agricultural, and socio-economic \cite{khan_wavelet_2020} \cite{gyaneshwar_contemporary_2023},  with each type influenced by critical climatic factors such as increased evaporation, transpiration, and insufficient precipitation. In this review, we will consider only meteorological drought. 

Meteorological Drought forecasting is a complex process that involves analyzing various climatic and environmental factors to anticipate future drought conditions. Statistical models and drought indices, such as the Standardized Precipitation Index (SPI) or Palmer Drought Severity Index (PDSI) \cite{pathak_comparison_2020}, are utilized to quantify drought severity and monitor changes over time. These methodologies contribute to a comprehensive understanding of drought dynamics and aid in the formulation of effective drought management and adaptation measures. However, they often face challenges in capturing the complexities of drought dynamics.

\begin{table*}[htp]
\caption{Deep Learning in Drought}\label{table_drought}
\centering
\begin{tabular}{@{}llll@{}}
\toprule
Task & Approach & Ref\\ 

\midrule

SPI Forecast & ANFIS, FFNN, MLR 
& \citep{bacanli_adaptive_2009}\\

& ARIMA, ANN, SVR, WA-ANN, WA-SVR
& \citep{belayneh_long-term_2014}\\

& ANN, SVR, WANN
& \citep{belayneh_short-term_2016}\\

& ANN, WANN, ARIMA, SARIMA
& \citep{djerbouai_drought_2016}\\

& ARIMA, ANN, WANN
& \citep{zhang_multi-models_2017}\\

& EMD-DBN
& \citep{agana_emd-based_2018}\\

& WP-ANN, WP-SVR
& \citep{das_hybrid_2020}\\

SWSI and SIAP Forecast & ANN
& \citep{khan_wavelet-ann_2018}\\

SPEI Forecast & ANFIS, hybrid WT-ARIMA-ANN
& \citep{soh_application_2018}\\

\bottomrule
\end{tabular}
\end{table*}

The study utilized ANFIS to forecast SPI-based drought indices using rainfall data from 10 stations in Central Anatolia, Turkey. The ANFIS architecture consists of five layers: input, rule, average, consequent, and output nodes. These models, named SPI-1, SPI-3, SPI-6, SPI-9, and SPI-12 for different time scales, are designed to capture diverse drought patterns by integrating SPI and precipitation data, addressing short-term, seasonal, and long-term variations. For each phase, a total of 20 distinct models with varied input combinations are developed. The evaluation includes performance metrics such as Root Mean Square Error (RMSE), Efficiency (E), and Correlation (CORR), and comparisons are made against Feed Forward Neural Networks (FFNN) and Multiple Linear Regression (MLR) models. Significantly, the ANFIS models exhibit exceptional performance, showcasing a notable ability to accurately identify dry and wet periods, particularly across extended time scales \cite{bacanli_adaptive_2009}.
In a different study, the focus is on long-term Standard Precipitation Index (SPI) drought forecasting (6 and 12 months lead time) in the Awash River Basin, Ethiopia. The study compares the efficacy of five data-driven models: ARIMA, ANN, SVR, WA-ANN, and WA-SVR. The ARIMA model involves three essential steps: model identification, parameterization, and validation. The significant lags are determined using ACF and PACF, guiding the selection of accurate and precise parameters. The ANN model employs a MLP structure with input, hidden, and output layers, trained using the Levenberg-Marquardt (LM) backpropagation algorithm. Lagged SPI values are utilized as input, with optimal input layer neurons determined via trial and error, while hidden layer neurons are selected using empirical methods. The SVR model employs a non-linear RBF kernel. The Wavelet Decomposition process encompasses CWT for time-frequency representation and DWT for efficient computation. The transformed time series serves as input for both ANN and SVR models. The performance of models is evaluated using RMSE, MAE, R2, and persistence. It is found that the WA-ANN model outperforms other models for forecasting SPI values over lead times of 6 and 12 months\cite{belayneh_long-term_2014}. 
In another study, the WA-ANN models outperform alternative approaches, providing the most accurate forecasts for SPI 3 and SPI 6 values over lead times of 1 and 3 months. This underscores the effectiveness of the WA-ANN architecture, characterized by 3 to 5 neurons in the input layer, 4 to 7 neurons in the hidden layer, and a single neuron in the output layer, contributing to improved short-term drought forecasting \cite{belayneh_short-term_2016}.
A comparative approach is employed to evaluate the performance of various forecasting models for drought using SPI as the indicator. Three primary models are investigated: ANN, WANN, and traditional stochastic models, namely ARIMA and Seasonal ARIMA (SARIMA). The focus is on SPI-3, SPI-6, and SPI-12 time scales, and the impact of wavelet preprocessing on model accuracy is explored for the Algerois Basin in North Algeria. ARIMA and Seasonal ARIMA models offer an empirical framework for modeling and predicting complex hydrologic systems, with nonseasonal ARIMA addressing stationary data and seasonal ARIMA handling nonstationarity through AR, MA operators, and differentiation parameters. An implementation of ANN-MLP involves a network structure comprising interlinked input, hidden, and output layers. The WA-ANN model utilizes Discrete Wavelet (DW) inputs derived from original SPI time series and corresponding un-decomposed SPI outputs, with a focus on assessing the impact of various mother wavelets to enhance model efficiency. The model performance is assessed using the Nash-Sutcliffe model efficiency coefficient (NSE), RSME, and MAE as evaluation metrics. The results demonstrate that the WANN model outperforms the ANN model for SPI-3 forecasts over up to six months while the SARIMA model shows satisfactory results for SPI-12 forecasts with a one-month lead time. However, all models experience reduced accuracy as lead times increase \cite{djerbouai_drought_2016}. 
In another study, three data-driven models, namely ARIMA, ANN, and WANN, are employed for drought forecasting based on the SPI at two time scales (SPI-6 and SPI-12) in the north of the Haihe River Basin. The effectiveness of the models is assessed using statistical tests like the Kolmogorov-Smirnov (K-S) test, Kendall rank correlation, and correlation coefficients (R2). ARIMA models with varying parameter combinations are employed, selecting the model that minimizes K–S distance and maximizes Kendall rank correlation. The chosen model's efficacy is validated through ACF and PACF plots. The ANN model underestimates certain instances of extreme drought or extreme precipitation, where the observed SPI values correspond to such extreme conditions. The WANN model outperforms other models, exhibiting higher correlation, lower K–S distance, and enhanced Kendall rank correlation. The comparison results show that the WANN model is the most suitable and effective for forecasting SPI-6 and SPI-12 values in the study area \cite{zhang_multi-models_2017}.
Furthermore, a hybrid predictive model is presented, combining EMD with a DBN for drought forecasting using the SSI across the Colorado River basin. DBN is constructed by stacking multiple RBM on top of each other, and the RBMs are trained using the contrastive divergence algorithm. EMD is utilized to decompose the data into IMF with varying frequencies. Some IMFs are found to contain noise or irrelevant information. A denoising technique is proposed, involving the use of Detrended Fluctuation Analysis (DFA) scaling exponents. A threshold (Hurst exponent 0.5) is applied to identify noisy IMFs, which are subsequently eliminated, and the relevant IMFs are aggregated for reconstruction. The DBN model, along with other models (MLP, SVR, EMD-MLP, EMD-SVR), is used to predict SSI-12 with lead times of one and two months. The evaluation metrics include RMSE, MAE, and NSE, and the EMD-DBN model outperforms all other models in the two-step ahead prediction \cite{agana_emd-based_2018}.  

The focus of the study is on drought forecasting with lead times of 1 month and 6 months for the Gulbarga district in Karnataka, India, using the SPI as the drought quantifying parameter. WPT is employed to preprocess SPI time series, generating decomposed coefficients used as inputs for ANN and SVR models. The SPI time series forecasting utilizes a BPNN with a 3-4-1 architecture. The network's weights and biases are determined using the gradient descent optimization algorithm, incorporating an adaptive learning rate of 0.45, momentum rate of 0.15, and 5000 learning cycles. The SVR utilizes a loss function based on Vapnik’s $\epsilon$-insensitive approach and incorporates the Gaussian radial basis function (RBF) kernel. This approach creates hybrid WP-ANN and WP-SVR models for drought forecasting, with Daubechies 4 (db4) wavelet chosen as the mother wavelet. It is observed that the hybrid WP-ANN model performs better than standalone approaches, with the forecast accuracy decreasing as the lead time increases \cite{das_hybrid_2020}.

In another study, an ANN model is used to forecast drought indices, including the Standardised Water Storage Index (SWSI) and the Standard Index of Annual Precipitation (SIAP). The dataset encompasses rainfall and water level information originating from the Langat River Basin in Malaysia, covering a time span of three decades (1986-2016). A feed-forward multilayer perceptron (MLP) structure is employed, comprising input, hidden, and output layers. This architecture is trained using the Levenberg–Marquardt (LM) back-propagation algorithm for both traditional artificial neural network (ANN) models and the wavelet-based artificial neural network (W-ANN) models. In the W-ANN approach, discrete wavelet transform (DWT) is applied to the input data, yielding subseries components. Subsequently, pertinent components are chosen from these subseries and integrated into the MLP to enhance the accuracy of forecasting. The outcomes demonstrate that the W-ANN model showcases improved performance, achieving heightened correlation coefficients \cite{khan_wavelet-ann_2018}. 

Another study employs two hybrid models, namely Wavelet-ARIMA-ANN (WAANN) and Wavelet-Adaptive Neuro-Fuzzy Inference System (WANFIS), to predict the Standardized Precipitation Evapotranspiration Index (SPEI) at the Langat River Basin for different time scales (1-month, 3-months, and 6-months). The input data are subjected to wavelet decomposition at a level of three, and the resulting components are employed as inputs for both ANN and ANFIS models. ANN models are constructed using Bayesian regularization backpropagation with a total of 1000 training epochs, and the optimal number of hidden neurons is identified through trial and error. The WANFIS involves normalizing decomposed historical SPEI series as input for Sugeno-type Fuzzy Inference System (FIS), chosen for its computational efficiency and compatibility with optimization techniques. This is followed by applying the ANFIS algorithm with determined training parameters to enhance model performance. It is found that the hybrid WT-ARIMA-ANN technique outperforms other models, providing better forecasts for both short-term and mid-term drought indices (SPEI 1, SPEI 3, and SPEI 6).\cite{soh_application_2018}.  

\subsection{Deep Learning in Heatwave and Cold waves}
Heatwaves are extreme weather event characterized by prolonged periods of excessively hot weather \cite{perkins_measurement_2013}, often accompanied by high humidity. During a heatwave, temperatures rise significantly above the average for a particular location and persist for an extended period, typically several days. Conversely, a coldwave is a meteorological phenomenon marked by a sudden and significant decrease in air temperature at the Earth's surface. This results in extremely low temperatures that can give rise to hazardous weather conditions, including frost formation and the formation of ice. Both events can have profound impacts on human health \cite{coumou_decade_2012}, \cite{lopez-bueno_effect_2021} particularly in vulnerable populations such as the elderly, children, and individuals with pre-existing health conditions, infrastructure, agriculture, ecosystems, and  can even result in mortality for human beings and livestocks. Additionally, heatwaves strain power grids due to increased air conditioning use, resulting in power outages, and can cause crop failures, wildfires and damage to infrastructure like roads, bridges, railways, and airports \cite{abdin_modeling_2019}.  The World Meteorological Organization's (WMO) annual report for 2023 highlighted the unprecedented heatwaves experienced in Europe during the summer, exacerbated by abnormally dry conditions. Tragically, these extreme heat events resulted in over 15,000 excess deaths across several countries, including Spain, Germany, the United Kingdom, France, and Portugal. These alarming findings underscore the pressing need for urgent and effective heatwave mitigation strategies and adaptive measures to safeguard vulnerable populations in the face of escalating climate challenges \cite{noauthor_wmo_2023}. 

Monitoring and predicting heatwaves and cold waves are crucial for preparedness \cite{lavaysse_predictability_2019} and mitigating potential risks, such as implementing appropriate measures to protect vulnerable populations and ensuring the efficient functioning of critical systems during extreme cold episodes. Both the events can be predicted using a range of approaches, including statistical models \cite{dodla_analysis_2017} and dynamic models such as GMC \cite{peng_toward_2011} \cite{nasim_future_2018} and RCM \cite{dosio_projection_2017} \cite{vautard_simulation_2013} \cite{singh_evaluation_2021} and machine learning techniques that analyze extensive datasets to identify patterns associated with heatwave occurrences. Deep learning can be utilized to extract features from various meteorological variables, such as temperature, humidity, wind patterns, and atmospheric pressure, to forecast the likelihood, frequency, duration and intensity of both heatwaves and coldwaves. 

\begin{table*}[htp]
\caption{Deep Learning in Heatwaves and Cold waves}\label{table_heatwave_coldwave}
\centering
\begin{tabular}{@{}llll@{}}
\toprule
Task & Approach & Ref\\ 
\midrule

Heatwave Forecast
& LSTM
& \citep{narkhede_empirical_2022}\\

& GNN
& \citep{li_regional_2023}\\

Heatwave and Cold wave Forecast
& ConvNet, CapsNets
& \citep{chattopadhyay_analog_2020}\\

SAT Forecast
& CNN, CNN-RP, CNN-RP-BIN
& \citep{fister_accurate_2023}\\

SAT and LST Forecast
& LSTM
& \citep{chung_correlation_2020}\\

\bottomrule
\end{tabular}
\end{table*}

The focus of the study is on heatwave monitoring and prediction, employing index-based monitoring, and LSTM-based prediction model in northern India up to 5-6 days ahead. The study employs IMD daily mean gridded surface temperature data (1951–2020) and NCMRWF-IMDAA reanalysis dataset for humidity and wind data (1979–2020). The objective of this study is to develop an operational framework that can monitor, track, and predict heatwaves in real-time over the Indian region, utilizing a combination of temperature indices, synoptic information, and an LSTM-based prediction model. The model's performance is evaluated using a correlation coefficient and root mean square error (RMSE). The results demonstrate that the proposed approach offers a promising approach to enhance heatwave preparedness and response strategies \cite{narkhede_empirical_2022}.
In another study, a GNN model is developed to predict regional summer heatwaves in North America. By utilizing daily weather data of 91 stations across CONUS and analyzing key meteorological variables, the model reduces computational burdens for immediate heatwave warnings and facilitates fast decision-making. The model utilizes an encoder-processor-decoder architecture for binary classification of heatwave events. Each node within the graph corresponds to a weather station, while the model employs a GAL – a form of nonlinear graph convolution. This GAL dynamically adjusts the adjacency matrix based on node features via attention mechanisms, thereby enhancing its expressiveness. The GAL computes specialized attention weights (AW) to capture interactions between nodes, encompassing influences from neighbors, to neighbors, and historical data. Moreover, the model's training incorporates a soft F1-score metric, effectively combining recall and precision to mitigate bias and maximize the F1-score. As a result, the GNN model achieves an impressive 90\% accuracy in predicting regional heatwave occurrences \cite{li_regional_2023}. 
Another study employs ConvNet and CapsNet to predict heatwaves and cold waves, predicting the occurrence and region of extreme weather patterns in North America. The study employs daily data from the Large-Ensemble Community Project (LENS) for surface air temperature (T2m) and geopotential height at 500 mb (Z500) during boreal summer and winter months from 1920 to 2005. The ConvNet architecture comprises 4 convolutional layers with ReLU activation, where the last two layers are followed by max-pooling (2x2, stride 1). The output feeds into a fully connected neural network with 200 neurons, featuring dropout regularization and L2 regularization to prevent overfitting. An adaptive learning rate is implemented through the ADAM optimizer, while a softmax layer assigns patterns to cluster indices based on the highest probability. On the other hand, CapsNet includes two convolutional layers with ReLU activation, followed by a primary capsule layer (eight capsules with eight convolution layers), utilizing the routing-by-agreement algorithm to convey information to a secondary capsule layer for cluster probability prediction. The squash function introduces nonlinearity, and a decoding layer with three fully connected layers aids pattern reconstruction. The framework's performance is evaluated using accuracy and recall metrics and compared against CNN and logistic regression. The CapsNet-based framework achieves notable accuracy and recall in predicting extreme temperature events based on Z500 patterns \cite{chattopadhyay_analog_2020}.

The challenge of long-term air temperature prediction in summer using AI techniques is addressed in the study. ECMWF's ERA5 reanalysis data spanning 1950 to 2021 for Paris (France) and Córdoba (Spain) is employed, incorporating nine vital meteorological variables, including 2m air temperature, sea surface temperature, wind components (10m and 100m), mean sea level pressure, soil water layer, and geopotential pressure. For each region, two experiments are carried out: one for shorter-term prediction and another for prolonged prediction time-horizons, therefore possibly indicating a heatwave or a coldwave occurrence. The research explores a diverse array of nine modeling approaches, encompassing Linear Regression (LR), Lasso Regression (Lasso), Polynomial Regression (Poly), AdaBoost, Decision Trees (DT), Random Forest (RF), Convolutional Neural Network (CNN), CNN with Recurrence Plots (RP+CNN), and RP+CNN with binary fusion (RP+CNN+BIN). Each method is assessed using MSE, MAE, Pearson and Spearman rank correlation coefficients, and optimal predictor variable subsets from exhaustive search. The CNN combined with RP approaches accurately detects maximum temperatures, indicating heatwaves, outperforming classical CNN and other machine learning techniques. \cite{fister_accurate_2023}. 
In another study, the investigation of the association between surface air temperature (SAT) and land surface temperature (LST) considering land use during heat and cold wave events is undertaken. The author employs LSTM with a memory block containing forget, input, and output gates. These gates utilize sigmoid layers and pointwise multiplication to govern the flow of data across the cell and neural networks, effectively managing data dynamics. The study uses Terra and Aqua MODIS daytime and nighttime LST data, along with observed air temperature data obtained from 79 weather stations under the Korea Meteorological Administration spanning the years 2008 to 2018. The performance of the model is conducted using metrics such as R-squared, Root Mean Square Error (RMSE), and Index of Agreement (IoA) \cite{chung_correlation_2020}.

\subsection{Deep Learning in Tropical Cyclone}
Tropical cyclones are low-pressure weather systems that form over warm tropical oceans between latitudes 23.5 degrees North and South, except in the South-Atlantic Ocean region \cite{verma2022adaptive}.  \\

\subsubsection{Frequency and Identification}
A study focuses on predicting TC frequency during the post-monsoon season based on large-scale climate variables such as geopotential height, relative humidity, sea-level pressure, and zonal wind. Three types of artificial neural networks, namely MLP, RBF, and GRNN, are employed to develop prediction models. The research methodology involves selecting significant predictors using correlation analysis and utilizing historical TC frequency data from 1971 to 2013. The models are trained with data from 1971 to 2002 and evaluated with independent data from 2003 to 2013. The MLP architecture consists of two hidden layers with five nodes in the first layer, three nodes in the second layer, and an output layer. The RBF network employs radial basis functions with optimized spread parameters of 0.6, while the GRNN employs a parallel structure with spread factors of 0.2. Results demonstrate that the MLP model outperforms RBF and GRNN models across various evaluation metrics, showing lower RMSE, higher correlation, and better agreement between predicted and observed TC counts \cite{nath_seasonal_2016}. 
Furthermore, a multistaged deep learning framework proposes incorporating a Mask R-CNN detector, a wind speed filter, and a classifier based on CNN to detect TCs. The Mask R-CNN detector with the R50 FPN model predicts TC locations, trained on RGB satellite images, and generates predictions with class labels, scores, segmentation masks, and bounding box coordinates. A Wind Speed Filter is applied to reduce false positives using a threshold of 34KT or higher. Cropped images based on bounding box coordinates from the detector are fed to the DenseNet169 CNN classifier to differentiate between true TCs and non-TCs. This methodology is optimized using Bayesian optimization techniques. The study uses Meteosat Visible Infra-Red Imager (MVIRI) IR satellite images from Meteosat 5 and Meteosat 7 in the Indian Ocean Data Coverage (IODC) region. The model is tested on a dataset of 171 images, including 88 TCs, indicating promising performance \cite{nair_deep_2022}. \\

\subsubsection{Genesis Forecast}
Traditionally, meteorologists relies on various physical models and statistical techniques to predict tropical cyclone (TC) genesis. However, these physical models often have limitations and simplifications, which can affect their accuracy in capturing the complex interactions and dynamics involved in TC genesis. On the other hand, statistical models have been used to analyze historical data and identify patterns and relationships between different meteorological variables and TC genesis. While statistical models can provide valuable insights and correlations, they may struggle to capture nonlinear and complex relationships present in the data. 

Short-term tropical cyclogenesis forecasting plays a critical role in predicting the formation and development of TCs within a relatively short time frame. A study investigated the detectability of TCs and their precursors using a CNN model across different basins, seasons, and lead times. The CNN architecture consists of four convolutional layers, three pooling layers, and three fully connected layers, and finally an output layer with two units for binary classification. The Adam optimizer is applied to the CNN to update the network parameters to minimize the loss function called binary cross-entropy. In the western North Pacific, the CNN successfully detects TCs and their precursors during the period of July to November, achieving high POD ranging from 79.0\% to 89.1\%, along with relatively low FAR ranging from 32.8\% to 53.4\%. Notably, the CNN exhibits impressive performance in detecting precursors, with detection results of 91.2\%, 77.8\%, and 74.8\% for precursors occurring 2, 5, and 7 days before their formation, respectively. This method displays promise for studying tropical cyclogenesis and exhibits robust performance even in regions with limited training data and short TC lifetimes. However, the detection of TCs and their precursors is found to be limited in cases where cloud cover is extremely small (< 30\%) or extremely large (> 95\%). Considering developing TCs and precursors as one category potentially affects the ability to detect pre-TCs. Additionally, model-specific biases are identified due to the CNN being trained solely on Nonhydrostatic ICosahedral Atmospheric Model (NICAM) dataset. Notably, the detection performance in the North Atlantic is relatively lower, which could be attributed to the scarcity of training data and shorter lifetimes of TCs in that particular region \cite{matsuoka_deep_2018}.

In the realm of long-term cyclogenesis forecasting, various approaches have been employed to improve the accuracy of predictions and provide insights into the behavior of TCs over an extended period. Taking a distinctive route, a study combines SOM and FFNN to investigate changes in TCs' GPI and its contributing factors for a global climate model. This study introduces a comprehensive methodology employing two types of artificial neural networks to project changes in North Atlantic tropical cyclone genesis potential under the warming climate of the twenty-first century. Through SOMs, archetypal patterns of GPI-related environmental variables are captured, arranging them on a two-dimensional grid to retain data topology. Concurrently, FBNNs identify the relative importance of these variables in driving projected GP changes. SOMs' training ensures the preservation of data relationships, while FBNNs calculate variable relevance for GP outcomes. The FBNNs' training involves conveying input signals to hidden-layer nodes, generating output via sigmoid functions. The neural network framework NEVPROP4 is employed for FBNN implementation. This dual-network approach yields significant insights into the intricate trends of TC genesis potential as they respond to evolving environmental conditions \cite{yip_application_2012}. \\

\begin{table*}[htp]
\caption{Deep Learning in Tropical Cyclone}\label{table_tc_genesis}
\centering
\begin{tabular}{@{}llll@{}}
\toprule
Task & Approach & Ref\\ 
\midrule

Frequency & MLP, RBF, GRNN
& \citep{nath_seasonal_2016}\\

Identification
& CNN, Mask R-CNN
& \citep{nair_deep_2022}\\

\midrule

Cyclogenesis Forecast
& CNN
& \citep{matsuoka_deep_2018}\\

& SOM-FNN
& \citep{yip_application_2012}\\

\midrule

Track Forecast
& HRBF
& \citep{lee_tropical_2000}\\

& ANN
& \citep{ali_predicting_2007}\\

& GAN
& \citep{ruttgers_prediction_2018}\\

& LSTM
& \citep{gao_nowcasting_2018}\\

& CNN
& \citep{giffard-roisin_tropical_2020}\\

\midrule

Intensity Forecast
& NN
& \citep{baik_neural_2000}\\

& MLP
& \citep{chaudhuri_intensity_2013}\\

& CNN
& \citep{pradhan_tropical_2018}\\

& RNN
& \citep{pan_tropical_2019}\\

& CNN
& \citep{zhang_tropical_2020}\\

& CNN
& \citep{tian_cnn-based_2020}\\

& double cascade CNN
& \citep{zhang_tropical_2021}\\

& CNN, VGG-19 
& \citep{devaraj_novel_2021}\\

& CNN
& \citep{zhuo_physics-augmented_2021}\\

& CNN
& \citep{wang_tropical_2022}\\

& CNN, VGG
& \citep{tan_tropical_2022}\\

RI Prediction
& CNN
& \citep{wei_investigating_2023}\\

& RNN, LSTM
& \citep{chen_deep_2023}\\

\bottomrule
\end{tabular}
\end{table*}

\subsubsection{Track Forecast}
The accurate prediction of TC tracks is crucial for effective disaster preparedness and response. In recent years, deep learning techniques have emerged as promising tools for improving TC track forecasting. In a study, researchers aimed to utilize the neural oscillatory elastic graph matching (NOEGM) technique for tropical cyclone (TC) pattern identification, and a hybrid radial basis function (HRBF) network integrated with time difference and structural learning (TDSL) algorithm for TC track prediction. The HRBF network employs three layers, with past network outputs introduced through time-delay units and influenced by a decay factor. The evaluation encompassed 120 TC cases spanning from 1985 to 1998. The NOEGM model achieved noteworthy results, with 98\% accurate segmentation and 97\% correct classification rates for TC pattern recognition. The HRBF model showcased an accuracy of over 86\% in TC track and intensity mining. In comparison with prevailing TC prediction models, the proposed approach demonstrated substantial enhancements, reducing forecast errors by more than 30\% and achieving a remarkable 14\% enhancement in 48-hour track forecast accuracy. \cite{lee_tropical_2000}. 
Utilizing an extensive dataset spanning 32 years of cyclone best track analysis, researchers constructed an ANN model to forecast TC positions 24 hours in advance. Notably, this model incorporates inputs from the two most recent 6-hourly positions, along with the present latitude and longitude, while predicting positions for a 24-hour lead time. Through a systematic exploration, a range of both linear and nonlinear transfer functions, such as Radial Basis Function and linear least squares optimization, are evaluated. Furthermore, different configurations of hidden layers and neurons are experimented with to optimize performance. The chosen linear neural network architecture, driven by a pseudo invert learning algorithm, yields remarkable results, achieving MAE as low as 0.75 degrees for latitude and 0.87 degrees for longitude. The model's effectiveness is reinforced by a comparison of average errors: the Limited Area Model (LAM), the National Centre for Environmental Prediction based Quasi Lagrangian Model (QLM), and ANN models exhibit errors of 132.6 km, 142.0 km, and 127.5 km, respectively. These findings underscore the potential accuracy of the ANN-based approach in cyclone tracking, particularly within a 24-hour prediction window \cite{ali_predicting_2007}. 
\citeauthor{ruttgers_prediction_2018} leveraged a Generative Adversarial Network (GAN) to anticipate the paths of typhoons. This was accomplished by merging satellite images from the KMA and reanalysis data from the ECMWF dataset, covering the time span from 1993 to 2017, with a focus on typhoons that could impact the Korean Peninsula. Training data comprised cropped segments of historical typhoon images, while full-scale images were employed for testing purposes. The GAN framework consisted of a generator, which utilized multi-scale capabilities to generate diverse images, and a discriminator to differentiate between authentic and generated images. Inputs encompassed meteorological variables such as Sea Surface Temperature (SST), Sea Pressure, Relative Humidity (RH), Surface velocity field (zonal and meridional components), Velocity field at 950 mb pressure level, and Vertical wind shear (at 850 mb and 200 mb pressure levels). The training process involved iteratively optimizing both networks using distinct loss functions: L2 loss for quantifying image disparities, gradient difference loss to amplify image clarity, and adversarial loss to challenge the discriminator's ability to distinguish real from generated images. The findings underscored the GAN's efficacy in predicting typhoon trajectories and cloud formations. Accuracy in predicting typhoon center positions was assessed, revealing that a majority fell within 80 km (65.5\%), a notable portion within 80-120 km (31.5\%), and a smaller fraction exceeded 120 km (3.0\%). The overall prediction error was significantly reduced to 67.2 km, compared to 95.6 km when relying solely on observational data. The GAN's ability to anticipate cloud movement patterns underscored its potential in capturing dynamic phenomena \cite{ruttgers_prediction_2018}. 
An algorithm based on LSTM is employed for desirable 6-24 hour nowcasting of typhoon tracks on historical typhoon data from 1949 to 2011 in China's Mainland. The model's architecture encompasses three layers: input, hidden, and output, featuring 20 LSTM cells in the hidden layer and 2 neurons in the output layer. Through backpropagation utilizing the BPTT algorithm, errors are minimized by comparing predictions with actual observed tracks, thus offering a substantial advancement in typhoon track prediction \cite{gao_nowcasting_2018}. 
An innovative approach is introduced to predict tropical cyclone movement over a 24-hour timeframe by combining historical trajectory data and reanalysis atmospheric images, particularly wind and pressure fields. The technique involves adopting a dynamic frame of reference that follows the storm center, thus enhancing the precision of forecasts. The model's versatility is demonstrated by its capability to rapidly provide forecasts for newly emerging storms, a crucial asset for real-time predictions. Leveraging an extensive database spanning more than three decades and over 3,000 storms, sampled at six-hour intervals, the approach integrates past displacement data, metadata, wind fields, and geopotential height fields to capture diverse information. The methodology involves separate training of the Wind CNN, Pressure CNN, and Past Tracks + Meta NN, followed by integration into a fused network. Training incorporates root mean square error (RMSE) as the loss function, with regularization to prevent overfitting. This fusion network not only enhances prediction accuracy but also significantly reduces testing time, making it a promising advancement for real-time forecasting in the realm of tropical cyclones \cite{giffard-roisin_tropical_2020}. \\

\subsubsection{Intensity Prediction}
Tropical cyclone intensity prediction is a critical aspect of forecasting, and various approaches have been explored in recent research. In a study, an advanced neural network model is employed to predict tropical cyclone intensity changes in the western North Pacific, incorporating climatology, persistence, and synoptic factors. The neural network architecture consists of three layers: an input layer with 11 units representing climatology, persistence, and synoptic predictors; a hidden layer with 11 units capturing complex relationships; and an output layer predicting intensity changes. The models analyzed include a multiple linear regression model with climatology and persistence predictors (R-CP), a neural network model with the same predictors (N-CP), a multiple linear regression model with climatology, persistence, and synoptic predictors (R-CPS), and a neural network model incorporating all predictors (N-CPS). The performance of these models is assessed through average intensity prediction errors across different prediction intervals. The N-CPS model demonstrates superior performance in predicting tropical cyclone intensity changes, especially over shorter time intervals, while the N-CP model shows slight superiority over the R-CPS model \cite{baik_neural_2000}. 
In a study, the use of MLP models capable of forecasting intensity changes at 3-hour intervals beyond 72 hours in the North Indian Ocean, specifically in the Bay of Bengal and Arabian Sea, is explored. The architecture of the MLP model incorporates central pressure (CP), maximum sustained wind speed (MSWS), pressure drop (PD), total ozone column (TOC), and sea surface temperature (SST) as inputs for predicting cyclone intensity. The model's effectiveness is assessed using metrics like RSME and MAE, revealing the MLP's superior performance compared to other models like RBFN, MLR, and OLR for forecasting cyclone intensity. The models' individual performances are evaluated for various cyclones, accounting for varying sea surface temperatures over the Arabian Sea and Bay of Bengal \cite{chaudhuri_intensity_2013}. 
In a study, a CNN model is used to estimate the intensity of tropical cyclones in the Atlantic and Pacific regions. The proposed model uses a comprehensive dataset comprising two distinct components: a collection of 48,828 infrared (IR) hurricane images sourced from the Marine Meteorology Division of the U.S. Naval Research Laboratory, and HURDAT2 data to label these images. The model's architecture integrates convolutional layers with varying filter sizes and strides, followed by strategic max-pooling for down-sampling. Complemented by local response normalization and fully connected layers with ReLU activation, the model incorporates regularization techniques like dropout to prevent overfitting. Weight updates are fine-tuned through SGD optimization with momentum, and a specialized Softmax loss layer facilitates accurate multi-class classification. By autonomously extracting pivotal features from TC images, this methodology achieves better accuracy and reduced RSME, indicating a significant advancement in tropical cyclone intensity estimation \cite{pradhan_tropical_2018}. 
Additionally, a study investigates the application of RNN for forecasting TC intensity by leveraging historical observation data in the Western North Pacific since 1949. The RNN architecture captures intricate relationships among sequential elements—longitude, latitude, and intensity—across input, hidden, and output layers. Integrating a backpropagation through time optimization algorithm, the model refines weights and biases. Employing a cross-entropy loss function, it gauges disparities between predicted and actual TC intensity, with the hidden layer employing the tanh activation function. Notably, the model excels, achieving a compelling 5.1 m/s error in 24-hour forecasts, outperforming select dynamical models and closely approximating subjective predictions.\cite{pan_tropical_2019}. 
A dual-branch CNN model is proposed for estimating tropical cyclone (TC) intensity in the Northwest Pacific. The model exhibits strong performance for tropical storm and super typhoon categories but demonstrates reduced accuracy for moderate intensity and the weakest tropical depression category. The architecture of the TCIENet model comprises two parallel CNN branches designed for processing infrared and water vapor images. Each branch includes essential modules for feature extraction, water vapor attention, and intensity regression, with the overall goal of capturing the intricate relationship between image patterns and TC intensity. The training is facilitated by the Adam optimizer, utilizing techniques such as Softmax operation, dropout regularization, and L1 and L2 loss functions to enhance its predictive capability. The research also delves into the impact of diverse image sizes and model components on intensity estimation accuracy, leveraging metrics like RMSE, MAE, bias, and absolute error to evaluate the model's effectiveness \cite{zhang_tropical_2020}. 
\citeauthor{tian_cnn-based_2020} presented a novel CNN-based hybrid model designed to accurately estimate tropical cyclone intensity by harnessing 46,919 infrared images sourced from the Pacific Northwest and Atlantic Ocean. This architecture incorporates a classification model, fine-grained regression models, and a Back-propagation neural network. The classification model effectively categorizes TC samples into distinct intensity levels, thereby guiding the selection of appropriate regression models. The model's optimization is carried out using the Adam optimization algorithm, while a cross-validation loss function is employed for both classification and regression tasks. Notably, the model achieves exceptional accuracy and remarkably low RSME, outperforming the existing methodologies \cite{tian_cnn-based_2020}. 
The TCICENet model offers a novel approach to accurately classifying and estimating tropical cyclone intensity using infrared satellite images from the northwest Pacific Ocean basin. This model adopts a cascading deep-CNN architecture consisting of two essential components: TC intensity classification (TCIC) and TC intensity estimation (TCIE). The TCIC module employs convolutional layers to categorize TC intensity into three specific classes, while the TCIE module, inspired by a modified AlexNet structure, predicts intensity values across different TC intensity categories. Notably, the TCIC module employs a cross-entropy loss with L2 regularization, and the TCIE module employs a SmoothL1 loss function for precise intensity estimation. The model's effectiveness is validated using a dataset encompassing 1001 TCs from 1981 to 2019, partitioned into distinct sets for training, validation, and testing. Evaluation based on intensity estimation metrics reveals impressive performance, achieving an overall root mean square error of 8.60 kt and a mean absolute error of 6.67 kt in comparison to best track data. \cite{zhang_tropical_2021}.
In a separate study, a CNN model is utilized to predict the intensity levels of hurricanes using IR satellite imagery data from HURSAT and wind speed data from the HURDAT2 of the Greater Houston region. The architecture involves sequential layers: input, convolution, pooling, and fully connected, guided by ReLU activations, MSE loss, and RmsProp/Adam optimizers. This facilitates accurate hurricane intensity estimation, pattern recognition for storm categorization by severity, achieving lower RMSE (7.6 knots) and MSE (6.68 knots) through batch normalization and dropout layers. Additionally, a VGG19 model is employed to evaluate the extent of damage and automate annotation of satellite imagery data. The VGG 19 model undergoes fine-tuning for hurricane damage prediction and classification of severe weather events. The optimization process is guided by the Adam Optimizer, utilizing MSE as the foundational loss function. The models are subjected to rigorous evaluation, encompassing a diverse set of metrics including RMSE, MAE, MSE, and Relative RSME. Notably, the model demonstrates remarkable performance, achieving a 98\% accuracy in predicting hurricane damage and a 97\% accuracy in classifying severe weather events \cite{devaraj_novel_2021}. 
Furthermore, a model called DeepTCNet is proposed specifically designed for TC intensity and size estimation in the North Atlantic by using IBTrACS and the Hurricane Satellite dataset. The study harnesses CNN as the core architecture within DeepTCNet to estimate TC intensity and wind radii from IR imagery. Extensive experimentation establishes VGGNet with 13 layers and compact (3 × 3) convolutional filters as the optimal configuration, forming the foundational structure for DeepTCNet. The evaluation presents MAE for TC intensity estimation (measured in knots) on the test dataset across various depths and kernel sizes in VGGNet's initial convolutional layer. Leveraging the Adam optimization with default parameters, learning occurs through the adoption of MAE as the loss function. This holistic approach exemplifies the seamless fusion of physics-augmented deep learning, culminating in enhanced TC analysis and prediction capabilities \cite{zhuo_physics-augmented_2021}. 
The study focuses on estimating TC intensity using a CNN model. Satellite IR imagery and Best Track data are employed to analyze 97 TC cases over the Northwest Pacific Ocean from 2015 to 2018. The CNN architecture encompasses an input layer, four convolutional layers, four pooling layers, two fully connected layers, and an output layer, resulting in the derivation of eight intensity values. Notably, the multicategory CNN achieves an accuracy of 84.8\% for TC intensity estimation, which further improves to 88.9\% through conversion to a binary classification task. \cite{wang_tropical_2022}. 
Another study proposes a CNN model for estimating TC intensity using Himawari-8 satellite cloud products, including cloud optical thickness (CLOT), cloud top temperature (CLTT), cloud top height (CLTH), cloud effective radius (CLER), and cloud type (CLTY). The model's architecture is based on the VGG framework, enhanced with attention mechanisms and residual learning to improve precision while reducing parameter count. The CNN comprises four convolutional blocks with progressively larger filter sizes, integrating residual learning at different levels and a Convolutional Block Attention Module (CBAM) after a maximum pooling layer. Batch normalization and dropout layers are employed to counter overfitting. The model is optimized using the Adam optimizer and MAE loss function. It undergoes training and tuning through six-fold cross-validation and is evaluated on independent test data, utilizing real-time typhoon track information from the western North Pacific basin alongside Himawari-8 cloud products \cite{tan_tropical_2022}. 
These studies contribute valuable insights into improving tropical cyclone intensity forecasts through the utilization of advanced neural network models.

In addition to general tropical cyclone intensity forecasting, a particularly challenging aspect is predicting Rapid Intensification (RI), where a tropical cyclone undergoes a sudden and significant strengthening over a short timeframe. RI is a critical phenomenon due to its potential to escalate a relatively mild storm into a highly destructive force, posing severe threats to coastal communities and infrastructure. 

To address the complexities of RI prediction, a CNN model called TCNET was developed to enhance the prediction of RI in tropical cyclones by extracting features from large-scale environmental conditions. The study used ECMWF ERA-Interim reanalysis data and the SHIPS database. TCNET's architecture consists of data filters, a customized sampler (GMM-SMOTE), an XGBoost classifier, and hyperparameter tuning. This model's performance outperforms COR-SHIPS and LLE-SHIPS in RI prediction, yielding superior results in terms of kappa, PSS, POD, and FAR metrics. Moreover, TCNET identifies previously unexplored variables, such as ozone mass mixing ratio, that influence RI. The training of TCNET involves backpropagation, utilizing mean square error as the loss function and Adam optimizer for weight updates of the filters \cite{wei_investigating_2023}. 
In another study, deep learning models including RNN and LSTM were explored for predicting tropical cyclone intensity and rapid intensification (RI). The proposed approach involved convolutional layers for autonomous feature extraction from satellite images, an RNN block with ConvLSTM cells for feature evolution, and a final output regressor composed of convolutional and dense layers to forecast tropical cyclone intensity (Vmax) at +24 hours. Additionally, the study introduced a deep learning ensemble strategy involving 20 models with diverse designs, effectively improving TC intensity and RI prediction by incorporating both conventional and satellite-derived features. This ensemble method offered intensity distributions for deterministic predictions, RI likelihood estimation, and prediction uncertainty assessment, yielding improved RI detection probabilities and reduced false-alarm rates compared to operational forecasts for western Pacific TCs \cite{chen_deep_2023}.

\section{Challenges}
\label{sec:challenges}
The effective utilization of DL models in weather forecasting is accompanied by several challenges that require careful consideration. In this section, we explore several key challenges associated with the application of deep learning in the field of weather forecasting.\\

\subsubsection{Data Availability}
Data availability is crucial for advancing the capabilities of deep learning models in meteorological applications. Limited access to historical records, real-time observations, and specialized data sources can hamper model development and evaluation \cite{chen2014big}. Addressing data availability challenges requires establishing robust data-sharing frameworks, promoting data collaboration between meteorological organizations, and exploring innovative approaches to gather and enhance meteorological data.\\

\subsubsection{Data Quality}
Ensuring data quality poses a significant challenge for deep learning models in meteorology. Weather data, obtained from various sources like weather stations, satellites, and radars, may have limitations in terms of spatial coverage, temporal resolution, and accuracy. Missing or inaccurate observations can introduce biases and errors. For example, inadequate temperature measurements in remote regions due to limited weather station distribution can lead to incomplete climate models, potentially affecting the accuracy of long-term weather predcitions.\\

\subsubsection{Model Architecture}
DL models often consist of complex architectures with numerous layers and parameters, posing challenges in their design and optimization for weather forecasting. Determining the optimal network architecture, selecting appropriate activation functions, and managing computational resources are critical task in developing efficient DL models for meteorological applications \cite{goodfellow2016deep}.\\

\subsubsection{Hybrid Approach}
Combining DL techniques with traditional physical models can leverage the strengths of both approaches, leading to more accurate and reliable predictions. DL models excel at learning complex patterns and capturing nonlinear relationships in large datasets \cite{chen2014big}, while traditional physical models provide valuable insights into the underlying physical processes. For instance, coupling a deep learning algorithm with a NWP model can allow the DL component to capture intricate spatial patterns in satellite imagery, while the physical model contributes its understanding of atmospheric physics. This collaborative approach offers the potential for more precise predictions of complex meteorological events. However, integrating DL models with existing frameworks triggers challenges such as resource requirement, data quality and interpretability. \\

\subsubsection{Data Heterogeneity}
Data heterogeneity refers to the diversity of data sources, formats, and features, which can complicate the integration and analysis of different data types. For instance, in the development of a deep learning model for weather prediction, information must be accumulated from various sources such as satellites, radar systems, automatic weather stations, numerical models and manual observations. However, each of these sources employs its own distinct method of storing data. Satellites often employ formats like TIFF or GeoTIFF, while radar data may utilize formats such as HDF5 or NetCDF, and other sources could have unique formats. To ensure the seamless operation of a weather prediction model, these different types of data must be harmonized, enabling the model to effectively learn from the combined data and resulting in accurate and reliable weather forecasts. \\

\subsubsection{Model Explainability}
Model explainibility refers to the how a model process and transform input into corresponding output, making the process transparent and easy to comprehend \cite{chakraborty2017interpretability}. For instance, if the model predicts upcoming rain, meteorologists need to comprehend the specific meteorological variables and potential biases influencing its predictions. This understanding becomes crucial as the model transitions to real-world application, allowing its developers to provide insights into its functioning.

\section{Discussion and Future Directions}
\label{sec:discussion}
The integration of deep learning methods into the study of extreme weather events brings about a significant transformation, expanding our understanding and predictive abilities. These advanced models not only excel in deciphering intricate spatial relationships but also stand out in unraveling the complex timing patterns inherent in meteorological phenomena. This advancement holds the potential to greatly improve weather forecasting accuracy across a wide range of events, from thunderstorms and lightning occurrences to the tracking of tropical cyclones.

At the heart of this innovation lies the natural capacity of deep learning models to identify and replicate non-linear relationships within the complex fabric of atmospheric data. By analyzing extensive sets of information, these models unearth hidden patterns and interactions that conventional methods struggle to capture. However, as we move forward with these promising advancements, it becomes crucial to address certain key challenges that must be overcome to fully realize the potential of deep learning in meteorology. One such challenge revolves around the continuous need for high-quality and comprehensive data. The effectiveness of deep learning models relies on their exposure to a diverse array of carefully curated data points. This emphasizes the importance of creating robust data pipelines and well-organized datasets. Furthermore, the inherently complex nature of deep learning architectures presents a dilemma regarding their interpretability. Ensuring that the decisions made by these intricate models can be comprehended and validated by experts in meteorology remains an ongoing endeavor.

The future of weather prediction calls for the exploration of ensemble techniques that combine the strengths of various models to produce more comprehensive and accurate forecasts. This pursuit involves developing innovative approaches that seamlessly integrate deep learning models with traditional numerical weather prediction methods, drawing on the well-established physical understanding of atmospheric processes. The convergence of deep learning capabilities with specialized insights in meteorology emerges as a fertile area for further exploration. Hybrid models that blend empirical meteorological knowledge with the computational power of deep learning offer a promising path to enhancing forecast accuracy and reinforcing our ability to handle the multifaceted impacts of extreme weather events.

In the broader context of this review, a clear message underscores the persistent drive for progress, necessitating ongoing collaboration and interdisciplinary synergy. By harnessing the capabilities of deep learning and pushing the boundaries of meteorological understanding, we are positioned to empower decision-makers and stakeholders with invaluable tools to proactively mitigate the far-reaching consequences of the ever-evolving realm of extreme weather events.

In weather forecasting, there exist several research gaps that need to be addressed to enhance the capabilities and effectiveness of these models. Firstly, in drought prediction, current studies lack long-term forecasting capabilities and are limited in spatial resolution. Improving these aspects is crucial to provide accurate and detailed information on drought conditions, enabling proactive mitigation measures. In the case of tropical cyclones, there is a notable absence of studies focused on pattern identification. Efforts should be directed towards reducing the cone of uncertainty by improving track accuracy, size estimation, and spatial distribution of cyclones. More research is needed in predicting RI and associated weather phenomena such as storm surge, floods, and quantitative precipitation forecasts. The lack of practical success stories in this area underscores the need for further investigation and advancements. In heatwave prediction, there is a paucity of research focused on the frequency and duration of heatwaves. The prediction of severe thunderstorms poses its own set of challenges. To improve forecasts in this domain, exploring other ensemble techniques, incorporating feature selection methods, and leveraging dynamic graph modeling approaches can be beneficial. Integrating data from multiple NWP models, along with high-resolution NWP models, holds promise for enhancing thunderstorm forecasts. The intensity, frequency, and location prediction of lightning strikes require further attention. Monitoring and predicting lightning strikes, especially in discreetly distributed scenarios, remain complex tasks that require advanced techniques and data integration. Radar and satellite data play a crucial role in weather forecasting. However, challenges persist in utilizing radar data to make predictions without clear indications of initial convections. Exploiting the early-stage signals of convections using radar and satellite data can aid in improving forecast accuracy, particularly in mitigating false alarms. Cloud-related weather forecasting also faces challenges, including high computation time and resource requirements. Inefficient observations due to rain, strong winds, foggy conditions, and sunsets further hinder the efficiency of cloud-related forecasting methods. Addressing these challenges is essential to unlock the full potential of deep learning in cloud prediction. Further research is required to enhance hail detection, size estimation, forecasting, and damage assessment methods using deep learning techniques, despite recent advancements in hail-related studies. Improving models, such as NWP, remains a substantial challenge. 

\section{Conclusion} 
\label{sec:conclusion}
This review highlighted the significant advancements and promising potential of deep learning techniques in the field of meteorology, specifically in extreme weather events. Deep learning models, such as CNN and RNN, demonstrated their effectiveness in various applications, including cyclone prediction, severe rainfall and hail prediction, cloud and snow detection, rainfall-induced flood, landslide forecasting, and more. The utilization of deep learning algorithms allowed researchers to extract intricate patterns and features from complex meteorological datasets, leading to improved accuracy and performance in weather prediction and analysis. These models showed remarkable skill in capturing spatial and temporal dependencies in weather data, enabling more accurate predictions of extreme events and enhancing our understanding of their underlying dynamics. Furthermore, deep learning methods offered advantages over traditional statistical approaches by automatically learning representations and hierarchies of features, eliminating the need for manual feature engineering. This allowed for more efficient and effective analysis of large-scale meteorological datasets, facilitating the development of advanced forecasting models and decision-support systems. Deep learning models also provided an alternative approach by directly learning the relationships between input observations and output variables from data, circumventing the computational bottlenecks and time lags associated with physics-based models. However, further advancements were needed to enhance the performance and efficiency of these models in weather forecasting applications. Closing these research gaps and advancing the field of deep learning in weather forecasting would contribute to more accurate, reliable, and timely predictions, ultimately benefiting various sectors and society as a whole.

\section*{Acknowledgment}
The author would like to express heartfelt gratitude to the India Meteorological Development (IMD) and the Indian Institute of Information Technology, Allahabad (IIIT Allahabad) for their invaluable support and contributions to this journal. Their guidance, resources, and assistance have been instrumental in the successful completion of this research.

\section*{Conflicts of Interest}
The authors declare no conflict of interest.

\section*{Abbreviations}
The following abbreviations are used in this manuscript:
\begin{table}[htp]
\begin{tabular}{ll}
ANFIS     & Adaptive Neuro-Fuzzy Inference System               \\ 
ANN       & Artificial Neural Networks                          \\
ARIMA     & Autoregressive Integrated Moving Average            \\
BPNN      & Back Propagation Neural Network                     \\
CAPE      & Convective Available Potential Energy               \\
CapsNets  & Capsule Neural Networks                             \\
CBAM      & Convolutional Block Attention Module                \\
CFS       & Climate Forecast System                             \\
CG        & Cloud-To-Ground                                     \\
CNN       & Convolutional Neural Networks                       \\
Conv2d    & 2D Convolution Layer                                \\
CRF       & Conditional Random Fields                           \\
CRPS      & Cumulative Ranked Probability Score                 \\
DBN       & Deep Belief Network                                 \\
DL        & Deep Learning                                       \\
DLNN      & Deep Learning Neural Network                        \\
DNN       & Deep Neural Network                                 \\
EEMD      & Ensemble Empirical Mode Decomposition               \\
EMD       & Empirical Mode Decomposition                        \\
ENSO      & El Niño-Southern Oscillation                        \\
FAR       & False Alarm Rate                                    \\
FFNN      & Feed Forward Neural Networks                        \\
GAN       & Generative Adversarial Network                      \\
GNN       & Graph Neural Network                                \\
GNSS      & Global Navigation Satellite System                  \\
GPI       & Genesis Potential Index                             \\
GRNN      & Generalized Regression Neural Network               \\
GRU       & Gated Recurrent Unit                                \\
HR        & Heavy Rain                                          \\
HREF      & High-Resolution Ensemble Forecast                   \\
IMD       & India Meteorological Department                     \\
IMF       & Intrinsic Mode Function                             \\
IR        & Infrared                                             \\
JTWC      & Joint Typhoon Warning Center                        \\
LM-ResNet & Lightning Monitoring Residual Network               \\
LRCN      & Long-Term Recurrent Convolutional Network           \\
LST       & Land Surface Temperature                            \\
LSTM      & Long Short-Term Memory                              \\
MAE       & Mean Absolute Error                                 \\
ML        & Machine Learning                                    \\
MLP       & Multi-Layer Perceptrons                             \\
MLR       & Multiple Linear Regression                          \\
MNLR      & Multivariate Non-Linear Regression                  \\
NARX      & Nonlinear Autoregressive Exogenous                  \\
NCEP      & National Centers For Environmental Prediction       \\
NIO       & North Indian Ocean                                  \\
NNGA      & Neural Network-Genetic Algorithm                    \\
NWP       & Numerical Weather Prediction                        \\
PCA       & Principal Component Analysis                        \\
PDSI      & Palmer Drought Severity Index                       \\
POD       & Probability Of Detection                            \\
POSH      & Probability Of Severe Hail                          \\
QPE       & Quantitative Precipitation Estimation               \\
RBF       & Radial Basis Function                               \\
RBM       & Restricted Boltzmann Machines                       \\
RF        & Random Forest                                       \\                                 
\end{tabular}
\end{table}
\begin{table}[h]
\begin{tabular}{ll}
RI        & Rapid Intensification                               \\
RMSE      & Root Mean Square Error                              \\
RNN       & Recurrent Neural Networks                           \\
RP        & Recurrence Plots                                    \\
SAT       & Surface Air Temperature                             \\
SCW       & Severe Convective Weather                           \\
SGD       & Stochastic Gradient Descent                         \\
SIAP      & Standard Index Of Annual Precipitation              \\
SOM       & Self-Organizing Maps                                \\
SPEI      & Standardized Precipitation Evapotranspiration Index \\
SPI       & Standardized Precipitation Index                    \\
SSI       & Standardized Streamflow Index                       \\
SVM       & Support Vector Machine                              \\
SWE       & Snow Water Equivalent                               \\
SWG       & Stochastic Weather Generator                        \\
SWSI      & Standardised Water Storage Index                    \\
TC        & Tropical Cyclone                                    \\
TPW       & Total Precipitable Water                            \\
VAEGAN    & Variational Autoencoder Gan                         \\
VGG       & Visual Geometry Group                               \\
WANN      & Wavelet-Artificial Neural Network                   \\
WGAN      & Wasserstein Generative Adversarial Network          \\
WMO       & World Meteorological Organization'S                 \\
WPT       & Wavelet Packet Transform                           
\end{tabular}
\end{table}

\bibliographystyle{IEEEtranN}
\bibliography{my_lib}

\begin{thebibliography}{95}
\providecommand{\natexlab}[1]{#1}
\providecommand{\url}[1]{#1}
\csname url@samestyle\endcsname
\providecommand{\newblock}{\relax}
\providecommand{\bibinfo}[2]{#2}
\providecommand{\BIBentrySTDinterwordspacing}{\spaceskip=0pt\relax}
\providecommand{\BIBentryALTinterwordstretchfactor}{4}
\providecommand{\BIBentryALTinterwordspacing}{\spaceskip=\fontdimen2\font plus
\BIBentryALTinterwordstretchfactor\fontdimen3\font minus
  \fontdimen4\font\relax}
\providecommand{\BIBforeignlanguage}[2]{{%
\expandafter\ifx\csname l@#1\endcsname\relax
\typeout{** WARNING: IEEEtranN.bst: No hyphenation pattern has been}%
\typeout{** loaded for the language `#1'. Using the pattern for}%
\typeout{** the default language instead.}%
\else
\language=\csname l@#1\endcsname
\fi
#2}}
\providecommand{\BIBdecl}{\relax}
\BIBdecl

\bibitem[WMO()]{wmoWorldMeteorological}
\BIBentryALTinterwordspacing
WMO, ``{W}orld {M}eteorological {O}rganization.'' [Online]. Available:
  \url{https://public.wmo.int/en}
\BIBentrySTDinterwordspacing

\bibitem[Easterling et~al.(2000)Easterling, Meehl, Parmesan, Changnon, Karl,
  and Mearns]{easterling2000climate}
D.~R. Easterling, G.~A. Meehl, C.~Parmesan, S.~A. Changnon, T.~R. Karl, and
  L.~O. Mearns, ``Climate extremes: observations, modeling, and impacts,''
  \emph{science}, vol. 289, no. 5487, pp. 2068--2074, 2000.

\bibitem[Beniston and Stephenson(2004)]{beniston2004extreme}
M.~Beniston and D.~B. Stephenson, ``Extreme climatic events and their evolution
  under changing climatic conditions,'' \emph{Global and planetary change},
  vol.~44, no. 1-4, pp. 1--9, 2004.

\bibitem[Alexander et~al.(2006)Alexander, Zhang, Peterson, Caesar, Gleason,
  Klein~Tank, Haylock, Collins, Trewin, Rahimzadeh,
  et~al.]{alexander2006global}
L.~V. Alexander, X.~Zhang, T.~C. Peterson, J.~Caesar, B.~Gleason,
  A.~Klein~Tank, M.~Haylock, D.~Collins, B.~Trewin, F.~Rahimzadeh
  \emph{et~al.}, ``Global observed changes in daily climate extremes of
  temperature and precipitation,'' \emph{Journal of Geophysical Research:
  Atmospheres}, vol. 111, no.~D5, 2006.

\bibitem[Tebaldi et~al.(2006)Tebaldi, Hayhoe, Arblaster, and
  Meehl]{tebaldi2006going}
C.~Tebaldi, K.~Hayhoe, J.~M. Arblaster, and G.~A. Meehl, ``Going to the
  extremes: an intercomparison of model-simulated historical and future changes
  in extreme events,'' \emph{Climatic change}, vol.~79, no. 3-4, pp. 185--211,
  2006.

\bibitem[IMD()]{imdHomeIndia}
IMD, ``India meteorological department,'' https://mausam.imd.gov.in/.

\bibitem[Espeholt et~al.(2022{\natexlab{a}})Espeholt, Agrawal, S{\o}nderby,
  Kumar, Heek, Bromberg, Gazen, Carver, Andrychowicz, Hickey,
  et~al.]{espeholt2022deep}
L.~Espeholt, S.~Agrawal, C.~S{\o}nderby, M.~Kumar, J.~Heek, C.~Bromberg,
  C.~Gazen, R.~Carver, M.~Andrychowicz, J.~Hickey \emph{et~al.}, ``Deep
  learning for twelve hour precipitation forecasts,'' \emph{Nature
  communications}, vol.~13, no.~1, pp. 1--10, 2022.

\bibitem[Goodfellow et~al.(2016)Goodfellow, Bengio, and
  Courville]{goodfellow2016deep}
I.~Goodfellow, Y.~Bengio, and A.~Courville, \emph{Deep learning}.\hskip 1em
  plus 0.5em minus 0.4em\relax MIT press, 2016.

\bibitem[Bauer et~al.(2015)Bauer, Thorpe, and Brunet]{bauer2015quiet}
P.~Bauer, A.~Thorpe, and G.~Brunet, ``The quiet revolution of numerical weather
  prediction,'' \emph{Nature}, vol. 525, no. 7567, pp. 47--55, 2015.

\bibitem[Trenberth(2011)]{trenberth2011changes}
K.~E. Trenberth, ``Changes in precipitation with climate change,''
  \emph{Climate research}, vol.~47, no. 1-2, pp. 123--138, 2011.

\bibitem[Abdin et~al.(2019)Abdin, Fang, and Zio]{abdin_modeling_2019}
I.~Abdin, Y.-P. Fang, and E.~Zio, ``A modeling and optimization framework for
  power systems design with operational flexibility and resilience against
  extreme heat waves and drought events,'' \emph{Renewable and Sustainable
  Energy Reviews}, vol. 112, pp. 706--719, 2019.

\bibitem[Coumou and Rahmstorf(2012)]{coumou_decade_2012}
D.~Coumou and S.~Rahmstorf, ``A decade of weather extremes,'' \emph{Nature
  climate change}, vol.~2, no.~7, pp. 491--496, 2012.

\bibitem[Heidari et~al.(2023)Heidari, Lin, Sarmiento, Toreti, and
  Xoplaki]{heidari_towards_2023}
F.~Heidari, Q.~Lin, E.~F.~E. Sarmiento, A.~Toreti, and E.~Xoplaki, ``Towards
  the development of an ai-based early warning system: a deep learning approach
  to bias correct and downscale seasonal climate forecasts,'' Copernicus
  Meetings, Tech. Rep., 2023.

\bibitem[Zhou et~al.(2019)Zhou, Zheng, Li, Dong, and
  Zhang]{zhou_forecasting_2019}
K.~Zhou, Y.~Zheng, B.~Li, W.~Dong, and X.~Zhang, ``Forecasting different types
  of convective weather: A deep learning approach,'' \emph{Journal of
  Meteorological Research}, vol.~33, pp. 797--809, 2019.

\bibitem[Vosper et~al.(2023{\natexlab{a}})Vosper, Watson, Harris, McRae,
  Santos-Rodriguez, Aitchison, and Mitchell]{vosper_deep_2023}
E.~Vosper, P.~Watson, L.~Harris, A.~McRae, R.~Santos-Rodriguez, L.~Aitchison,
  and D.~Mitchell, ``Deep learning for downscaling tropical cyclone rainfall to
  hazard-relevant spatial scales,'' \emph{Journal of Geophysical Research:
  Atmospheres}, p. e2022JD038163, 2023.

\bibitem[Vosper et~al.(2023{\natexlab{b}})Vosper, Watson, Harris, McRae,
  Santos-Rodriguez, Aitchison, and Mitchell]{vosper2023deep}
------, ``Deep learning for downscaling tropical cyclone rainfall,'' Copernicus
  Meetings, Tech. Rep., 2023.

\bibitem[Mondini et~al.(2023)Mondini, Guzzetti, and Melillo]{mondini_deep_2023}
A.~C. Mondini, F.~Guzzetti, and M.~Melillo, ``Deep learning forecast of
  rainfall-induced shallow landslides,'' \emph{Nature communications}, vol.~14,
  no.~1, p. 2466, 2023.

\bibitem[Azad et~al.(2021)Azad, Islam, Rahman, and Ayen]{azad_development_2021}
M.~A.~K. Azad, A.~R. M.~T. Islam, M.~S. Rahman, and K.~Ayen, ``Development of
  novel hybrid machine learning models for monthly thunderstorm frequency
  prediction over bangladesh,'' \emph{Natural Hazards}, vol. 108, pp.
  1109--1135, 2021.

\bibitem[Guastavino et~al.(2022)Guastavino, Piana, Tizzi, Cassola, Iengo,
  Sacchetti, Solazzo, and Benvenuto]{guastavino_prediction_2022}
S.~Guastavino, M.~Piana, M.~Tizzi, F.~Cassola, A.~Iengo, D.~Sacchetti,
  E.~Solazzo, and F.~Benvenuto, ``Prediction of severe thunderstorm events with
  ensemble deep learning and radar data,'' \emph{Scientific Reports}, vol.~12,
  no.~1, p. 20049, 2022.

\bibitem[Essa et~al.(2022)Essa, Hunt, Gijben, and Ajoodha]{essa_deep_2022}
Y.~Essa, H.~G. Hunt, M.~Gijben, and R.~Ajoodha, ``Deep learning prediction of
  thunderstorm severity using remote sensing weather data,'' \emph{IEEE Journal
  of Selected Topics in Applied Earth Observations and Remote Sensing},
  vol.~15, pp. 4004--4013, 2022.

\bibitem[Lin et~al.(2019)Lin, Li, Geng, Jiang, Xu, Zheng, Yao, Lyu, and
  Zhang]{lin_attention-based_2019}
T.~Lin, Q.~Li, Y.-A. Geng, L.~Jiang, L.~Xu, D.~Zheng, W.~Yao, W.~Lyu, and
  Y.~Zhang, ``Attention-based dual-source spatiotemporal neural network for
  lightning forecast,'' \emph{IEEE Access}, vol.~7, pp. 158\,296--158\,307,
  2019.

\bibitem[Lu et~al.(2022)Lu, Zhang, Chen, Yu, and Wang]{lu_monitoring_2022}
M.~Lu, Y.~Zhang, M.~Chen, M.~Yu, and M.~Wang, ``Monitoring lightning location
  based on deep learning combined with multisource spatial data,'' \emph{Remote
  Sensing}, vol.~14, no.~9, p. 2200, 2022.

\bibitem[Qian et~al.(2022)Qian, Wang, Shi, Yao, Hu, Yang, and
  Ni]{qian_lightning_2022}
Z.~Qian, D.~Wang, X.~Shi, J.~Yao, L.~Hu, H.~Yang, and Y.~Ni, ``Lightning
  identification method based on deep learning,'' \emph{Atmosphere}, vol.~13,
  no.~12, p. 2112, 2022.

\bibitem[Yao et~al.(2022{\natexlab{a}})Yao, Chen, Thompson, and
  Cifelli]{yao_improved_2022}
S.~Yao, H.~Chen, E.~J. Thompson, and R.~Cifelli, ``An improved deep learning
  model for high-impact weather nowcasting,'' \emph{IEEE Journal of Selected
  Topics in Applied Earth Observations and Remote Sensing}, vol.~15, pp.
  7400--7413, 2022.

\bibitem[Espeholt et~al.(2022{\natexlab{b}})Espeholt, Agrawal, S{\o}nderby,
  Kumar, Heek, Bromberg, Gazen, Carver, Andrychowicz, Hickey,
  et~al.]{espeholt_deep_2022}
L.~Espeholt, S.~Agrawal, C.~S{\o}nderby, M.~Kumar, J.~Heek, C.~Bromberg,
  C.~Gazen, R.~Carver, M.~Andrychowicz, J.~Hickey \emph{et~al.}, ``Deep
  learning for twelve hour precipitation forecasts,'' \emph{Nature
  communications}, vol.~13, no.~1, pp. 1--10, 2022.

\bibitem[Nan et~al.(2023)Nan, Chen, Ding, Li, and Chen]{nan_deep_2023}
T.~Nan, J.~Chen, Z.~Ding, W.~Li, and H.~Chen, ``Deep learning-based
  multi-source precipitation merging for the tibetan plateau,'' \emph{Science
  China Earth Sciences}, vol.~66, no.~4, pp. 852--870, 2023.

\bibitem[Li et~al.(2023{\natexlab{a}})Li, Chen, and Han]{li_polarimetric_2023}
W.~Li, H.~Chen, and L.~Han, ``Polarimetric radar quantitative precipitation
  estimation using deep convolutional neural networks,'' \emph{IEEE
  Transactions on Geoscience and Remote Sensing}, 2023.

\bibitem[Lee et~al.(2022)Lee, Ahn, and Lee]{lee_incremental_2022}
Y.~Lee, M.-H. Ahn, and S.-J. Lee, ``Incremental learning with neural network
  algorithm for the monitoring pre-convective environments using geostationary
  imager,'' \emph{Remote Sensing}, vol.~14, no.~2, p. 387, 2022.

\bibitem[Pullman et~al.(2019)Pullman, Gurung, Maskey, Ramachandran, and
  Christopher]{pullman_applying_2019}
M.~Pullman, I.~Gurung, M.~Maskey, R.~Ramachandran, and S.~A. Christopher,
  ``Applying deep learning to hail detection: A case study,'' \emph{IEEE
  Transactions on Geoscience and Remote Sensing}, vol.~57, no.~12, pp.
  10\,218--10\,225, 2019.

\bibitem[Pulukool et~al.(2020)Pulukool, Li, and Liu]{pulukool_using_2020}
F.~Pulukool, L.~Li, and C.~Liu, ``Using deep learning and machine learning
  methods to diagnose hailstorms in large-scale thermodynamic environments,''
  \emph{Sustainability}, vol.~12, no.~24, p. 10499, 2020.

\bibitem[Gagne~II et~al.(2019)Gagne~II, Haupt, Nychka, and
  Thompson]{ii_interpretable_2019}
D.~J. Gagne~II, S.~E. Haupt, D.~W. Nychka, and G.~Thompson, ``Interpretable
  deep learning for spatial analysis of severe hailstorms,'' \emph{Monthly
  Weather Review}, vol. 147, no.~8, pp. 2827--2845, 2019.

\bibitem[Wu et~al.(2021)Wu, Shou, Ma, Lu, and Wang]{wu_estimation_2022}
Q.~Wu, Y.-X. Shou, L.-M. Ma, Q.~Lu, and R.~Wang, ``Estimation of maximum hail
  diameters from fy-4a satellite data with a machine learning method,''
  \emph{Remote Sensing}, vol.~14, no.~1, p.~73, 2021.

\bibitem[Wang et~al.(2022{\natexlab{a}})Wang, Fan, Tu, Li, and
  Chen]{wang_cloud_2022}
Z.~Wang, B.~Fan, Z.~Tu, H.~Li, and D.~Chen, ``Cloud and snow identification
  based on deeplab v3+ and crf combined model for gf-1 wfv images,''
  \emph{Remote Sensing}, vol.~14, no.~19, p. 4880, 2022.

\bibitem[Zhan et~al.(2017)Zhan, Wang, Shi, Cheng, Yao, and
  Sun]{zhan_distinguishing_2017}
Y.~Zhan, J.~Wang, J.~Shi, G.~Cheng, L.~Yao, and W.~Sun, ``Distinguishing cloud
  and snow in satellite images via deep convolutional network,'' \emph{IEEE
  geoscience and remote sensing letters}, vol.~14, no.~10, pp. 1785--1789,
  2017.

\bibitem[Yin et~al.(2022)Yin, Wang, Ni, and Hao]{yin_cloud_2022}
M.~Yin, P.~Wang, C.~Ni, and W.~Hao, ``Cloud and snow detection of remote
  sensing images based on improved unet3+,'' \emph{Scientific Reports},
  vol.~12, no.~1, p. 14415, 2022.

\bibitem[Sood et~al.(2022)Sood, Tiwari, Singh, Kaur, and
  Parida]{sood_glacier_2022}
V.~Sood, R.~K. Tiwari, S.~Singh, R.~Kaur, and B.~R. Parida, ``Glacier boundary
  mapping using deep learning classification over bara shigri glacier in
  western himalayas,'' \emph{Sustainability}, vol.~14, no.~20, p. 13485, 2022.

\bibitem[Wang et~al.(2022{\natexlab{b}})Wang, Su, Zhai, Meng, and
  Liu]{wang_snow_2022}
Y.~Wang, J.~Su, X.~Zhai, F.~Meng, and C.~Liu, ``Snow coverage mapping by
  learning from sentinel-2 satellite multispectral images via machine learning
  algorithms,'' \emph{Remote Sensing}, vol.~14, no.~3, p. 782, 2022.

\bibitem[Wang et~al.(2021{\natexlab{a}})Wang, Zhang, Wang, He, and
  Luo]{wang_automated_2021}
H.~Wang, L.~Zhang, L.~Wang, J.~He, and H.~Luo, ``An automated snow mapper
  powered by machine learning,'' \emph{Remote Sensing}, vol.~13, no.~23, p.
  4826, 2021.

\bibitem[Zhu et~al.(2021)Zhu, Zhang, Wang, Tian, Liu, Ma, Kan, and
  Chu]{zhu_downscaling_2021}
L.~Zhu, Y.~Zhang, J.~Wang, W.~Tian, Q.~Liu, G.~Ma, X.~Kan, and Y.~Chu,
  ``Downscaling snow depth mapping by fusion of microwave and optical
  remote-sensing data based on deep learning,'' \emph{Remote Sensing}, vol.~13,
  no.~4, p. 584, 2021.

\bibitem[Xing et~al.(2022)Xing, Hou, Huang, and Zhang]{xing_estimation_2022}
D.~Xing, J.~Hou, C.~Huang, and W.~Zhang, ``Estimation of snow depth from amsr2
  and modis data based on deep residual learning network,'' \emph{Remote
  Sensing}, vol.~14, no.~20, p. 5089, 2022.

\bibitem[Yao et~al.(2022{\natexlab{b}})Yao, Zhang, Jiang, Ewe, and
  Ng]{yao_snow_2022}
H.~Yao, Y.~Zhang, L.~Jiang, H.~T. Ewe, and M.~Ng, ``Snow parameters inversion
  from passive microwave remote sensing measurements by deep convolutional
  neural networks,'' \emph{Sensors}, vol.~22, no.~13, p. 4769, 2022.

\bibitem[Ghanjkhanlo et~al.(2020)Ghanjkhanlo, Vafakhah, Zeinivand, and
  Fathzadeh]{ghanjkhanlo_prediction_2020}
H.~Ghanjkhanlo, M.~Vafakhah, H.~Zeinivand, and A.~Fathzadeh, ``Prediction of
  snow water equivalent using artificial neural network and adaptive
  neuro-fuzzy inference system with two sampling schemes in semi-arid region of
  iran,'' \emph{Journal of Mountain Science}, vol.~17, no.~7, pp. 1712--1723,
  2020.

\bibitem[Marofi et~al.(2011)Marofi, Tabari, and
  Abyaneh]{marofi_predicting_2011}
S.~Marofi, H.~Tabari, and H.~Z. Abyaneh, ``Predicting spatial distribution of
  snow water equivalent using multivariate non-linear regression and
  computational intelligence methods,'' \emph{Water resources management},
  vol.~25, pp. 1417--1435, 2011.

\bibitem[Khan et~al.(2020)Khan, Muhammad, and El-Shafie]{khan_wavelet_2020}
M.~M.~H. Khan, N.~S. Muhammad, and A.~El-Shafie, ``Wavelet based hybrid
  ann-arima models for meteorological drought forecasting,'' \emph{Journal of
  Hydrology}, vol. 590, p. 125380, 2020.

\bibitem[Gyaneshwar et~al.(2023)Gyaneshwar, Mishra, Chadha, Raj~Vincent,
  Rajinikanth, Pattukandan~Ganapathy, and
  Srinivasan]{gyaneshwar_contemporary_2023}
A.~Gyaneshwar, A.~Mishra, U.~Chadha, P.~D. Raj~Vincent, V.~Rajinikanth,
  G.~Pattukandan~Ganapathy, and K.~Srinivasan, ``A contemporary review on deep
  learning models for drought prediction,'' \emph{Sustainability}, vol.~15,
  no.~7, p. 6160, 2023.

\bibitem[Pathak and Dodamani(2020)]{pathak_comparison_2020}
A.~A. Pathak and B.~Dodamani, ``Comparison of meteorological drought indices
  for different climatic regions of an indian river basin,'' \emph{Asia-Pacific
  Journal of Atmospheric Sciences}, vol.~56, pp. 563--576, 2020.

\bibitem[Bacanli et~al.(2009)Bacanli, Firat, and Dikbas]{bacanli_adaptive_2009}
U.~G. Bacanli, M.~Firat, and F.~Dikbas, ``Adaptive neuro-fuzzy inference system
  for drought forecasting,'' \emph{Stochastic Environmental Research and Risk
  Assessment}, vol.~23, pp. 1143--1154, 2009.

\bibitem[Belayneh et~al.(2014)Belayneh, Adamowski, Khalil, and
  Ozga-Zielinski]{belayneh_long-term_2014}
A.~Belayneh, J.~Adamowski, B.~Khalil, and B.~Ozga-Zielinski, ``Long-term spi
  drought forecasting in the awash river basin in ethiopia using wavelet neural
  network and wavelet support vector regression models,'' \emph{Journal of
  Hydrology}, vol. 508, pp. 418--429, 2014.

\bibitem[Belayneh et~al.(2016)Belayneh, Adamowski, and
  Khalil]{belayneh_short-term_2016}
A.~Belayneh, J.~Adamowski, and B.~Khalil, ``Short-term spi drought forecasting
  in the awash river basin in ethiopia using wavelet transforms and machine
  learning methods,'' \emph{Sustainable Water Resources Management}, vol.~2,
  pp. 87--101, 2016.

\bibitem[Djerbouai and Souag-Gamane(2016)]{djerbouai_drought_2016}
S.~Djerbouai and D.~Souag-Gamane, ``Drought forecasting using neural networks,
  wavelet neural networks, and stochastic models: case of the algerois basin in
  north algeria,'' \emph{Water Resources Management}, vol.~30, pp. 2445--2464,
  2016.

\bibitem[Zhang et~al.(2017)Zhang, Li, Chen, Pu, and
  Xiang]{zhang_multi-models_2017}
Y.~Zhang, W.~Li, Q.~Chen, X.~Pu, and L.~Xiang, ``Multi-models for spi drought
  forecasting in the north of haihe river basin, china,'' \emph{Stochastic
  environmental research and risk assessment}, vol.~31, pp. 2471--2481, 2017.

\bibitem[Agana and Homaifar(2018)]{agana_emd-based_2018}
N.~A. Agana and A.~Homaifar, ``Emd-based predictive deep belief network for
  time series prediction: An application to drought forecasting,''
  \emph{Hydrology}, vol.~5, no.~1, p.~18, 2018.

\bibitem[Das et~al.(2020)Das, Naganna, Deka, and Pushparaj]{das_hybrid_2020}
P.~Das, S.~R. Naganna, P.~C. Deka, and J.~Pushparaj, ``Hybrid wavelet packet
  machine learning approaches for drought modeling,'' \emph{Environmental Earth
  Sciences}, vol.~79, pp. 1--18, 2020.

\bibitem[Khan et~al.(2018)Khan, Muhammad, and El-Shafie]{khan_wavelet-ann_2018}
M.~M.~H. Khan, N.~S. Muhammad, and A.~El-Shafie, ``Wavelet-ann versus ann-based
  model for hydrometeorological drought forecasting,'' \emph{Water}, vol.~10,
  no.~8, p. 998, 2018.

\bibitem[Soh et~al.(2018)Soh, Koo, Huang, and Fung]{soh_application_2018}
Y.~Soh, C.~H. Koo, Y.~Huang, and K.~Fung, ``Application of artificial
  intelligence models for the prediction of standardized precipitation
  evapotranspiration index (spei) at langat river basin, malaysia,''
  \emph{Computers and electronics in agriculture}, vol. 144, pp. 164--173,
  2018.

\bibitem[Perkins and Alexander(2013)]{perkins_measurement_2013}
S.~E. Perkins and L.~V. Alexander, ``On the measurement of heat waves,''
  \emph{Journal of climate}, vol.~26, no.~13, pp. 4500--4517, 2013.

\bibitem[L{\'o}pez-Bueno et~al.(2021)L{\'o}pez-Bueno, Navas-Mart{\'\i}n,
  D{\'\i}az, Mir{\'o}n, Luna, S{\'a}nchez-Mart{\'\i}nez, Culqui, and
  Linares]{lopez-bueno_effect_2021}
J.~A. L{\'o}pez-Bueno, M.~{\'A}. Navas-Mart{\'\i}n, J.~D{\'\i}az, I.~J.
  Mir{\'o}n, M.~Y. Luna, G.~S{\'a}nchez-Mart{\'\i}nez, D.~Culqui, and
  C.~Linares, ``The effect of cold waves on mortality in urban and rural areas
  of madrid,'' \emph{Environmental Sciences Europe}, vol.~33, no.~1, pp. 1--14,
  2021.

\bibitem[WMO(2023)]{noauthor_wmo_2023}
WMO, ``Wmo annual report highlights continuous advance of climate change,'' WMO
  Annual Report, Tech. Rep., 2023.

\bibitem[Lavaysse et~al.(2019)Lavaysse, Naumann, Alfieri, Salamon, and
  Vogt]{lavaysse_predictability_2019}
C.~Lavaysse, G.~Naumann, L.~Alfieri, P.~Salamon, and J.~Vogt, ``Predictability
  of the european heat and cold waves,'' \emph{Climate Dynamics}, vol.~52, pp.
  2481--2495, 2019.

\bibitem[Dodla et~al.(2017)Dodla, Satyanarayana, and
  Desamsetti]{dodla_analysis_2017}
V.~B. Dodla, G.~C. Satyanarayana, and S.~Desamsetti, ``Analysis and prediction
  of a catastrophic indian coastal heat wave of 2015,'' \emph{Natural Hazards},
  vol.~87, pp. 395--414, 2017.

\bibitem[Peng et~al.(2011)Peng, Bobb, Tebaldi, McDaniel, Bell, and
  Dominici]{peng_toward_2011}
R.~D. Peng, J.~F. Bobb, C.~Tebaldi, L.~McDaniel, M.~L. Bell, and F.~Dominici,
  ``Toward a quantitative estimate of future heat wave mortality under global
  climate change,'' \emph{Environmental health perspectives}, vol. 119, no.~5,
  pp. 701--706, 2011.

\bibitem[Nasim et~al.(2018)Nasim, Amin, Fahad, Awais, Khan, Mubeen, Wahid,
  Rehman, Ihsan, Ahmad, et~al.]{nasim_future_2018}
W.~Nasim, A.~Amin, S.~Fahad, M.~Awais, N.~Khan, M.~Mubeen, A.~Wahid, M.~H.
  Rehman, M.~Z. Ihsan, S.~Ahmad \emph{et~al.}, ``Future risk assessment by
  estimating historical heat wave trends with projected heat accumulation using
  simclim climate model in pakistan,'' \emph{Atmospheric Research}, vol. 205,
  pp. 118--133, 2018.

\bibitem[Dosio(2017)]{dosio_projection_2017}
A.~Dosio, ``Projection of temperature and heat waves for africa with an
  ensemble of cordex regional climate models,'' \emph{Climate Dynamics},
  vol.~49, no. 1-2, pp. 493--519, 2017.

\bibitem[Vautard et~al.(2013)Vautard, Gobiet, Jacob, Belda, Colette,
  D{\'e}qu{\'e}, Fern{\'a}ndez, Garc{\'\i}a-D{\'\i}ez, Goergen, G{\"u}ttler,
  et~al.]{vautard_simulation_2013}
R.~Vautard, A.~Gobiet, D.~Jacob, M.~Belda, A.~Colette, M.~D{\'e}qu{\'e},
  J.~Fern{\'a}ndez, M.~Garc{\'\i}a-D{\'\i}ez, K.~Goergen, I.~G{\"u}ttler
  \emph{et~al.}, ``The simulation of european heat waves from an ensemble of
  regional climate models within the euro-cordex project,'' \emph{Climate
  Dynamics}, vol.~41, pp. 2555--2575, 2013.

\bibitem[Singh et~al.(2021)Singh, Mall, Dadich, Verma, Singh, and
  Gupta]{singh_evaluation_2021}
S.~Singh, R.~Mall, J.~Dadich, S.~Verma, J.~Singh, and A.~Gupta, ``Evaluation of
  cordex-south asia regional climate models for heat wave simulations over
  india,'' \emph{Atmospheric Research}, vol. 248, p. 105228, 2021.

\bibitem[Narkhede et~al.(2022)Narkhede, Chattopadhyay, Lekshmi, Guhathakurta,
  Kumar, and Mohapatra]{narkhede_empirical_2022}
N.~Narkhede, R.~Chattopadhyay, S.~Lekshmi, P.~Guhathakurta, N.~Kumar, and
  M.~Mohapatra, ``An empirical model-based framework for operational monitoring
  and prediction of heatwaves based on temperature data,'' \emph{Modeling Earth
  Systems and Environment}, vol.~8, no.~4, pp. 5665--5682, 2022.

\bibitem[Li et~al.(2023{\natexlab{b}})Li, Yu, Huang, Wang, and
  Sharma]{li_regional_2023}
P.~Li, Y.~Yu, D.~Huang, Z.-H. Wang, and A.~Sharma, ``Regional heatwave
  prediction using graph neural network and weather station data,''
  \emph{Geophysical Research Letters}, vol.~50, no.~7, p. e2023GL103405, 2023.

\bibitem[Chattopadhyay et~al.(2020)Chattopadhyay, Nabizadeh, and
  Hassanzadeh]{chattopadhyay_analog_2020}
A.~Chattopadhyay, E.~Nabizadeh, and P.~Hassanzadeh, ``Analog forecasting of
  extreme-causing weather patterns using deep learning,'' \emph{Journal of
  Advances in Modeling Earth Systems}, vol.~12, no.~2, p. e2019MS001958, 2020.

\bibitem[Fister et~al.(2023)Fister, P{\'e}rez-Aracil,
  Pel{\'a}ez-Rodr{\'\i}guez, Del~Ser, and Salcedo-Sanz]{fister_accurate_2023}
D.~Fister, J.~P{\'e}rez-Aracil, C.~Pel{\'a}ez-Rodr{\'\i}guez, J.~Del~Ser, and
  S.~Salcedo-Sanz, ``Accurate long-term air temperature prediction with machine
  learning models and data reduction techniques,'' \emph{Applied Soft
  Computing}, vol. 136, p. 110118, 2023.

\bibitem[Chung et~al.(2020)Chung, Lee, Jang, Lee, and
  Kim]{chung_correlation_2020}
J.~Chung, Y.~Lee, W.~Jang, S.~Lee, and S.~Kim, ``Correlation analysis between
  air temperature and modis land surface temperature and prediction of air
  temperature using tensorflow long short-term memory for the period of
  occurrence of cold and heat waves,'' \emph{Remote Sensing}, vol.~12, no.~19,
  p. 3231, 2020.

\bibitem[Verma et~al.(2022)Verma, Agarwal, and Srivastava]{verma2022adaptive}
S.~Verma, A.~Agarwal, and K.~Srivastava, ``An adaptive approach to detect and
  track the cyclone path using remote sensing data,'' in \emph{2022 IEEE 19th
  India Council International Conference (INDICON)}.\hskip 1em plus 0.5em minus
  0.4em\relax IEEE, 2022, pp. 1--4.

\bibitem[Nath et~al.(2016)Nath, Kotal, and Kundu]{nath_seasonal_2016}
S.~Nath, S.~Kotal, and P.~Kundu, ``Seasonal prediction of tropical cyclone
  activity over the north indian ocean using three artificial neural
  networks,'' \emph{Meteorology and Atmospheric Physics}, vol. 128, pp.
  751--762, 2016.

\bibitem[Nair et~al.(2021)Nair, Srujan, Kulkarni, Alwadhi, Jain, Kodamana,
  Sandeep, and John]{nair_deep_2022}
A.~Nair, K.~S. Srujan, S.~R. Kulkarni, K.~Alwadhi, N.~Jain, H.~Kodamana,
  S.~Sandeep, and V.~O. John, ``A deep learning framework for the detection of
  tropical cyclones from satellite images,'' \emph{IEEE Geoscience and Remote
  Sensing Letters}, vol.~19, pp. 1--5, 2021.

\bibitem[Matsuoka et~al.(2018)Matsuoka, Nakano, Sugiyama, and
  Uchida]{matsuoka_deep_2018}
D.~Matsuoka, M.~Nakano, D.~Sugiyama, and S.~Uchida, ``Deep learning approach
  for detecting tropical cyclones and their precursors in the simulation by a
  cloud-resolving global nonhydrostatic atmospheric model,'' \emph{Progress in
  Earth and Planetary Science}, vol.~5, no.~1, pp. 1--16, 2018.

\bibitem[Yip and Yau(2012)]{yip_application_2012}
Z.~K. Yip and M.~Yau, ``Application of artificial neural networks on north
  atlantic tropical cyclogenesis potential index in climate change,''
  \emph{Journal of Atmospheric and Oceanic Technology}, vol.~29, no.~9, pp.
  1202--1220, 2012.

\bibitem[Lee and Liu(2000)]{lee_tropical_2000}
R.~S. Lee and J.~N. Liu, ``Tropical cyclone identification and tracking system
  using integrated neural oscillatory elastic graph matching and hybrid rbf
  network track mining techniques,'' \emph{IEEE Transactions on Neural
  Networks}, vol.~11, no.~3, pp. 680--689, 2000.

\bibitem[Ali et~al.(2007)Ali, Kishtawal, and Jain]{ali_predicting_2007}
M.~Ali, C.~Kishtawal, and S.~Jain, ``Predicting cyclone tracks in the north
  indian ocean: An artificial neural network approach,'' \emph{Geophysical
  research letters}, vol.~34, no.~4, 2007.

\bibitem[R{\"u}ttgers et~al.(2018)R{\"u}ttgers, Lee, and
  You]{ruttgers_prediction_2018}
M.~R{\"u}ttgers, S.~Lee, and D.~You, ``Prediction of typhoon tracks using a
  generative adversarial network with observational and meteorological data,''
  \emph{arXiv preprint arXiv:1812.01943}, 2018.

\bibitem[Gao et~al.(2018)Gao, Zhao, Pan, Li, Zhou, Xu, Zhong, and
  Shi]{gao_nowcasting_2018}
S.~Gao, P.~Zhao, B.~Pan, Y.~Li, M.~Zhou, J.~Xu, S.~Zhong, and Z.~Shi, ``A
  nowcasting model for the prediction of typhoon tracks based on a long short
  term memory neural network,'' \emph{Acta Oceanologica Sinica}, vol.~37, pp.
  8--12, 2018.

\bibitem[Giffard-Roisin et~al.(2020)Giffard-Roisin, Yang, Charpiat,
  Kumler~Bonfanti, K{\'e}gl, and Monteleoni]{giffard-roisin_tropical_2020}
S.~Giffard-Roisin, M.~Yang, G.~Charpiat, C.~Kumler~Bonfanti, B.~K{\'e}gl, and
  C.~Monteleoni, ``Tropical cyclone track forecasting using fused deep learning
  from aligned reanalysis data,'' \emph{Frontiers in big Data}, p.~1, 2020.

\bibitem[Baik and Paek(2000)]{baik_neural_2000}
J.-J. Baik and J.-S. Paek, ``A neural network model for predicting typhoon
  intensity,'' \emph{Journal of the Meteorological Society of Japan. Ser. II},
  vol.~78, no.~6, pp. 857--869, 2000.

\bibitem[Chaudhuri et~al.(2013)Chaudhuri, Dutta, Goswami, and
  Middey]{chaudhuri_intensity_2013}
S.~Chaudhuri, D.~Dutta, S.~Goswami, and A.~Middey, ``Intensity forecast of
  tropical cyclones over north indian ocean using multilayer perceptron model:
  Skill and performance verification,'' \emph{Natural Hazards}, vol.~65, pp.
  97--113, 2013.

\bibitem[Pradhan et~al.(2017)Pradhan, Aygun, Maskey, Ramachandran, and
  Cecil]{pradhan_tropical_2018}
R.~Pradhan, R.~S. Aygun, M.~Maskey, R.~Ramachandran, and D.~J. Cecil,
  ``Tropical cyclone intensity estimation using a deep convolutional neural
  network,'' \emph{IEEE Transactions on Image Processing}, vol.~27, no.~2, pp.
  692--702, 2017.

\bibitem[Pan et~al.(2019)Pan, Xu, and Shi]{pan_tropical_2019}
B.~Pan, X.~Xu, and Z.~Shi, ``Tropical cyclone intensity prediction based on
  recurrent neural networks,'' \emph{Electronics Letters}, vol.~55, no.~7, pp.
  413--415, 2019.

\bibitem[Zhang et~al.(2019)Zhang, Liu, and Hang]{zhang_tropical_2020}
R.~Zhang, Q.~Liu, and R.~Hang, ``Tropical cyclone intensity estimation using
  two-branch convolutional neural network from infrared and water vapor
  images,'' \emph{IEEE Transactions on Geoscience and Remote Sensing}, vol.~58,
  no.~1, pp. 586--597, 2019.

\bibitem[Tian et~al.(2020)Tian, Huang, Yi, Wu, and Wang]{tian_cnn-based_2020}
W.~Tian, W.~Huang, L.~Yi, L.~Wu, and C.~Wang, ``A cnn-based hybrid model for
  tropical cyclone intensity estimation in meteorological industry,''
  \emph{IEEE Access}, vol.~8, pp. 59\,158--59\,168, 2020.

\bibitem[Zhang et~al.(2021)Zhang, Wang, Ma, and Lu]{zhang_tropical_2021}
C.-J. Zhang, X.-J. Wang, L.-M. Ma, and X.-Q. Lu, ``Tropical cyclone intensity
  classification and estimation using infrared satellite images with deep
  learning,'' \emph{IEEE Journal of Selected Topics in Applied Earth
  Observations and Remote Sensing}, vol.~14, pp. 2070--2086, 2021.

\bibitem[Devaraj et~al.(2021)Devaraj, Ganesan, Elavarasan, and
  Subramaniam]{devaraj_novel_2021}
J.~Devaraj, S.~Ganesan, R.~M. Elavarasan, and U.~Subramaniam, ``A novel deep
  learning based model for tropical intensity estimation and post-disaster
  management of hurricanes,'' \emph{Applied Sciences}, vol.~11, no.~9, p. 4129,
  2021.

\bibitem[Zhuo and Tan(2021)]{zhuo_physics-augmented_2021}
J.-Y. Zhuo and Z.-M. Tan, ``Physics-augmented deep learning to improve tropical
  cyclone intensity and size estimation from satellite imagery,'' \emph{Monthly
  Weather Review}, vol. 149, no.~7, pp. 2097--2113, 2021.

\bibitem[Wang et~al.(2021{\natexlab{b}})Wang, Zheng, Li, Xu, Liu, and
  Zhang]{wang_tropical_2022}
C.~Wang, G.~Zheng, X.~Li, Q.~Xu, B.~Liu, and J.~Zhang, ``Tropical cyclone
  intensity estimation from geostationary satellite imagery using deep
  convolutional neural networks,'' \emph{IEEE Transactions on Geoscience and
  Remote Sensing}, vol.~60, pp. 1--16, 2021.

\bibitem[Tan et~al.(2022)Tan, Yang, Hu, Huang, and Chen]{tan_tropical_2022}
J.~Tan, Q.~Yang, J.~Hu, Q.~Huang, and S.~Chen, ``Tropical cyclone intensity
  estimation using himawari-8 satellite cloud products and deep learning,''
  \emph{Remote Sensing}, vol.~14, no.~4, p. 812, 2022.

\bibitem[Wei et~al.(2023)Wei, Yang, and Sun]{wei_investigating_2023}
Y.~Wei, R.~Yang, and D.~Sun, ``Investigating tropical cyclone rapid
  intensification with an advanced artificial intelligence system and gridded
  reanalysis data,'' \emph{Atmosphere}, vol.~14, no.~2, p. 195, 2023.

\bibitem[Chen et~al.(2023)Chen, Kuo, and Huang]{chen_deep_2023}
B.-F. Chen, Y.-T. Kuo, and T.-S. Huang, ``A deep learning ensemble approach for
  predicting tropical cyclone rapid intensification,'' \emph{Atmospheric
  Science Letters}, vol.~24, no.~5, p. e1151, 2023.

\bibitem[Chen and Lin(2014)]{chen2014big}
X.-W. Chen and X.~Lin, ``Big data deep learning: challenges and perspectives,''
  \emph{IEEE access}, vol.~2, pp. 514--525, 2014.

\bibitem[Chakraborty et~al.(2017)Chakraborty, Tomsett, Raghavendra, Harborne,
  Alzantot, Cerutti, Srivastava, Preece, Julier, Rao,
  et~al.]{chakraborty2017interpretability}
S.~Chakraborty, R.~Tomsett, R.~Raghavendra, D.~Harborne, M.~Alzantot,
  F.~Cerutti, M.~Srivastava, A.~Preece, S.~Julier, R.~M. Rao \emph{et~al.},
  ``Interpretability of deep learning models: A survey of results,'' in
  \emph{2017 IEEE smartworld, ubiquitous intelligence \& computing, advanced \&
  trusted computed, scalable computing \& communications, cloud \& big data
  computing, Internet of people and smart city innovation
  (smartworld/SCALCOM/UIC/ATC/CBDcom/IOP/SCI)}.\hskip 1em plus 0.5em minus
  0.4em\relax IEEE, 2017, pp. 1--6.

\end{thebibliography}

\end{document}